\documentclass[12pt]{iopart}
\usepackage{graphicx,cite}
\usepackage{iopams}
\newtheorem{thm}{Theorem}[section]
\newtheorem{defn}{Definition}[section]
\newtheorem{lem}{Lemma}[section]
\newtheorem{remk}{Remark}[section]
\newenvironment{prf}[1][Proof]{\noindent\textbf{#1:} }{\ \rule{0.5em}{0.5em}}
\newcommand{\bequ}{\begin{equation}}
\newcommand{\beqn}{\begin{eqnarray}}
\newcommand{\eequ}{\end{equation}}
\newcommand{\eeqn}{\end{eqnarray}}

\begin{document}
\title{Geometric theory on the elasticity of bio-membranes}
\author{Z C Tu\dag\ddag\footnote[7]{Present address: Computational Material Science
Center, National Institute for Materials Science, Tsukuba
305-0047, Japan. Email: tu.zhanchun@nims.go.jp} and Z C
Ou-Yang\dag\S}
\address{\dag Institute of
Theoretical Physics,
 Chinese Academy of Sciences,
 P. O. Box 2735 Beijing 100080, China}
 \address{\ddag Graduate School,
 Chinese Academy of Sciences, China}
\address{\S Center for Advanced Study,
 Tsinghua University, Beijing 100084, China}

\begin{abstract}
The purpose of this paper is to study the shapes and stabilities
of bio-membranes within the framework of exterior differential
forms. After a brief review of the current status in theoretical
and experimental studies on the shapes of bio-membranes, a
geometric scheme is proposed to discuss the shape equation of
closed lipid bilayers, the shape equation and boundary conditions
of open lipid bilayers and two-component membranes, the shape
equation and in-plane strain equations of cell membranes with
cross-linking structures, and the stabilities of closed lipid
bilayers and cell membranes. The key point of this scheme is to
deal with the variational problems on the surfaces embedded in
three-dimensional Euclidean space by using exterior differential
forms.
\end{abstract}
\maketitle

\section{Introduction}
Cell membranes play crucial role in living movements. They consist
of lipids, proteins and carbohydrates etc. There are many
simplified models for cell membranes in history \cite{Edidin}.
Among them, the widely accepted one is the fluid mosaic model
proposed by Singer and Nicolson in 1972 \cite{nicolson}. In this
model, a cell membrane is considered as a lipid bilayer where
lipid molecules can move freely in the membrane surface like
fluid, while proteins are embedded in the lipid bilayer. This
model suggests that the shape of the cell membrane is determined
by its lipid bilayer. Usually, the thickness of lipid bilayer is
about 4 nanometers which is much less than the scale of the cell
(about several micrometers). Therefore, we can use a geometrical
surface to describe the lipid bilayer.

In 1973, Helfrich \cite{Helfrich} proposed the curvature energy
per unit area of the bilayer \bequ\label{Helfrich}
f_c=(k_c/2)(2H+c_0)^2+\bar{k}K, \eequ where $k_c$ and $\bar{k}$
are elastic constants; and $H$, $K$, $c_0$ are the mean, Gaussian,
and spontaneous curvatures of the membrane surface, respectively.
We can safely ignore the thermodynamic fluctuation of the curved
bilayer at the room temperature because of $k_c\approx
10^{-19}J\gg k_BT$ \cite{Duwe,Mutz2}, where $k_B$ is the Boltzmann
factor and $T$ the room temperature. Based on Helfrich's curvature
energy, the free energy of the closed bilayer under the osmotic
pressure $p$ (the outer pressure minus the inner one) is written
as \bequ\label{free-e-closed} \mathcal{F}_H=\int (f_c+\mu)
dA+p\int dV, \eequ where $dA$ is the area element, $\mu$ the
surface tension of the bilayer, and $V$ the volume enclosed within
the lipid bilayer. Starting with above free energy, many
researchers studied the shapes of bilayers \cite{oy1,Reinhard}.
Especially, by taking the first order variation of the free
energy, Ou-Yang \emph{et al.} derived an equation to describe the
equilibrium shape of the bilayer \cite{oy2}:
\bequ\label{shape-closed} p-2\mu
H+k_c(2H+c_0)(2H^2-c_0H-2K)+k_c\nabla^2(2H)=0. \eequ They also
obtained that the threshold pressure for instability of spherical
bilayer was $p_{c}\sim k_c/R^3$, where $R$ being the radius of
spherical bilayer.

Recently, opening-up process of lipid bilayers by talin was
observed by Saitoh \textit{et al.} \cite{Hotani,Hotani2}, which
arose the interest of studying the shape equation and boundary
conditions of lipid bilayers with free exposed edges. Capovilla
\textit{et al.} first gave the shape equation and boundary
conditions \cite{Capovilla} of open lipid bilayers. They also
discussed the mechanical meaning of these equations
\cite{Capovilla,Capovilla2}. In recent paper, we also derived the
shape equation and boundary conditions in different way---using
exterior differential forms to deal with the variational problems
on curved surfaces \cite{tzc1}. It is necessary to further develop
this method because we have seen that it is much more concise than
the tensor method in recent book \cite{oy1} by one of the authors.

In fact, the structures of cell membranes are far more complex
than the fluid mosaic model. The cross-linking structures exist in
cell membranes where filaments of membrane skeleton link to
proteins mosaicked in lipid bilayers \cite{Chiras}. It is worth
discussing whether the cross-linking structures have effect on the
shapes and stabilities of cell membranes.

In the following contents, both lipid bilayers and cell membranes
are called bio-membranes. We will fully develop our geometric
method to study the shapes and stabilities of bio-membranes. Our
method might not new for mathematicians who are familiar with the
work by Griffiths and Bryant \textit{et
al.}\cite{Griffiths,Bryant}. Our method focuses on the application
aspect, but the work by Griffiths and Bryant \textit{et al.}
emphasizes on the geometric meaning. Otherwise, we notice a nice
review paper by Kamien \cite{Kamien}, where he give an
introduction to the classic differential geometry in soft
materials. Here we will show that exterior differential forms not
mentioned by Kamien might also be useful in the study of
bio-membranes. This paper is organized as follows: In
Sec.\ref{Math-pre}, we briefly introduce the basic concepts in
differential geometry and the variational theory of surfaces. In
Sec.\ref{close-bilayer}, we deal with variational problems on a
closed surface, and derive the shape equation of closed lipid
bilayers, and then discuss the mechanical stabilities of spherical
bilayers. In Sec.\ref{open-bilayer}, we deal with variational
problems on an open surface, and then derive the shape equation
and boundary conditions of open lipid bilayers as well as
two-component lipid bilayers. In Sec.\ref{cell-membrane}, we
derive the free energy of the cross-linking structure by analogy
with the theory of rubber elasticity, and regard the free energy
of the cell membrane as the sum of the free energy of the closed
bilayer and that of cross-linking structure. The shape equation,
in-plane stain equations, and mechanical stabilities of cell
membranes are discussed by taking the first and second order
variations of the total free energy. In Sec.\ref{conclusion}, we
summarize the new results obtained in this paper.

\section{Mathematical preliminaries\label{Math-pre}}
Here we assume that the readers are familiar with the basic
concepts in differential geometry, such as manifold, differential
form and Stokes theorem (see also \ref{appdiff}).

\subsection{Surfaces in three-dimensional Euclidean space, moving frame method}
\begin{figure}[!htp]
\begin{center}
\includegraphics[width=8cm]{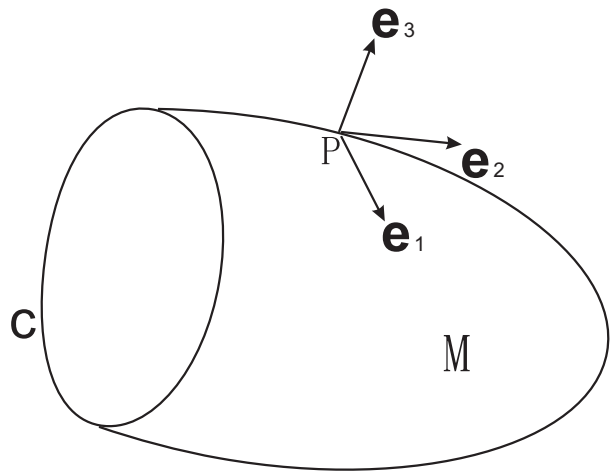}
\caption{\label{surfacem} A smooth and orientable surface $M$ with
an edge $C$.}\end{center}
\end{figure}

At every point $P$ of a smooth
and orientable surface $M$ in three-dimensional Euclidean space
$\mathbb{E}^3$, as shown in Fig.\ref{surfacem}, we can construct
an orthogonal system ${\mathbf{e}_1,\mathbf{e}_2,\mathbf{e}_3}$
such that $\mathbf{e}_3$ is the normal of the surface and
$\mathbf{e}_i\cdot\mathbf{e}_j=\delta_{ij},(i,j=1,2,3)$. We call
$\{P;\mathbf{e}_1,\mathbf{e}_2,\mathbf{e}_3\}$ a moving frame. For
the point in curve $C$, we let $\mathbf{e}_1$ be its tangent
vector and $\mathbf{e}_2$ point to the inner point of $M$. The
difference between two frames at point $P$ and $P'$ (which is very
close to $P$) is denoted by
\beqn\label{infiniter}d\mathbf{r}&=&\lim_{P\rightarrow P'}\overrightarrow{PP'}=\omega_1\mathbf{e}_1+\omega_2\mathbf{e}_2,\\
\label{dei}d\mathbf{e}_i&=&\omega_{ij}\mathbf{e}_j\quad (i=1,2,3),
\eeqn where $\omega_1$, $\omega_2$ and $\omega_{ij}$ $(i,j=1,2,3)$
are 1-forms.

It is easy to obtain $\omega_{ij}=-\omega_{ji}$ from
$\mathbf{e}_i\cdot\mathbf{e}_j=\delta_{ij}$. Additionally, using
$dd\mathbf{r}=0$ and $dd\mathbf{e}_i=0$, we obtain the structure
equations of the surface: \beqn \label{domega1}
d\omega_1=\omega_{12}\wedge\omega_2;\\
\label{domega2}
d\omega_2=\omega_{21}\wedge\omega_1;\\
\label{cardan}
\omega_1\wedge\omega_{13}+\omega_2\wedge\omega_{23}=0;\\
\label{domgaij} d\omega_{ij}=\omega_{ik}\wedge\omega_{kj}\quad
(i,j=1,2,3). \eeqn

If we considering the Cartan Lemma, (\ref{cardan}) suggests

\bequ  \label{omega13} \omega_{13}=a\omega_{1}+b\omega_{2}\quad
\textrm{and}\quad \omega_{23}=b\omega_{1}+c\omega_{2}. \eequ Thus
we have \cite{chen2}: \beqn
&&\textrm{The area element:\ }dA=\omega_1\wedge\omega_2, \\
&&\textrm{The first fundamental form:\ }I=d\mathbf{r}\cdot d\mathbf{r}%
=\omega_1^2+\omega_2^2,\nonumber\\
&&\textrm{The second fundamental form:\ }II=-d\mathbf{r}\cdot
d\mathbf{e}
_3=a\omega_1^2+2b\omega_1 \omega_2+c\omega_2^2,\nonumber\\
&&\textrm{The third fundamental form:\ }III=d\mathbf{e}_3\cdot d\mathbf{e}_3%
=\omega_{31}^2+\omega_{32}^2,\nonumber\\
&&\textrm{Mean curvature:\ }H=(a+c)/2 \label{meanh},\\
&&\textrm{Gaussian curvature:\ }K=ac-b^2\label{gassianK}. \eeqn

\subsection{Hodge star $\ast$ and Gauss mapping}
\subsubsection{Hodge star $\ast$}\hspace{1cm}
Here we just show the basic properties of Hodge star $\ast$ on
surface $M$. So as to its general definition, please see
Ref.\cite{westenholz}.

If $g,h$ are functions defined on 2D smooth surface $M$, then the
following formulas are valid: \beqn\label{starf}
\ast f=f\omega_1\wedge\omega_2;\\
\label{stardf} \ast \omega_1=\omega_2, \ast\omega_2=-\omega_1;
\\
\label{lap} d\ast df=\nabla^2 f\omega_1\wedge\omega_2, \eeqn where
$\nabla^2$ is Laplace-Beltrami operator.

It is easy to obtain the second Green identity
\begin{equation}\label{stoke2}
\int_M(fd\ast dh-hd\ast df)=\int_{\partial M}(f\ast dh-h\ast df)
\end{equation} through
Stokes theorem and the integration by parts. It follows that
\begin{equation}  \label{stoke3}
\int_Mfd\ast dh=\int_Mhd\ast df,
\end{equation}
if $M$ is a closed surface.

\subsubsection{Gauss mapping\label{gaussmapping}}
\hspace{1cm} The Gauss mapping $\mathcal{G}:M\rightarrow S^2$ is
defined as $\mathcal{G}(\mathbf{r})=\mathbf{e}_3(\mathbf{r})$,
where $S^2$ is a unit sphere. It induces a linear mapping
$\mathcal{G}^{\star}:\Lambda^1\rightarrow\Lambda^1$ such that:

(i) \ $\mathcal{G}^{\star}\omega_1=\omega_{13}$,
$\mathcal{G}^{\star}\omega_2=\omega_{23}$;

(ii) \ if $df=f_1\omega_{1}+f_2\omega_{2}$, then
$\mathcal{G}^{\star}df=f_1\mathcal{G}^{\star}\omega_1+f_2\mathcal{G}^{\star}\omega_2$.

Thus we can define a new differential operator
$\tilde{d}=\mathcal{G}^{\star}d$. Obviously, if
$df=f_1\omega_{1}+f_2\omega_{2}$, then
$\tilde{d}f=f_1\omega_{13}+f_2\omega_{23}$. If define a new
operator $\tilde{\ast}$ such that
$\tilde{\ast}\omega_{13}=\omega_{23}$ and
$\tilde{\ast}\omega_{23}=-\omega_{13}$, we have {\lem
$\int_M(fd\tilde{\ast}\tilde{d}h-hd\tilde{\ast}\tilde{d}f)=\int_{\partial
M}(f\tilde{\ast} \tilde{d}h-h\tilde{\ast} \tilde{d}f)$ for the
smooth functions $f$ and $h$ on $M$.\label{lemma1}}

 {\prf Using the integration by parts and Stokes theorem, we obtain
\beqn \int_M fd\tilde{\ast}\tilde{d}h=\int_{\partial
M}f\tilde{\ast}\tilde{d}h-\int_Mdf\wedge\tilde{\ast}\tilde{d}h,\label{midequ1}\\
\int_M hd\tilde{\ast}\tilde{d}f=\int_{\partial
M}h\tilde{\ast}\tilde{d}f-\int_Mdh\wedge\tilde{\ast}\tilde{d}f.\label{midequ2}
\eeqn

Otherwise, if let $df=f_1\omega_1+f_2\omega_2$ and
$dh=h_1\omega_1+h_2\omega_2$, we can prove
$df\wedge\tilde{\ast}\tilde{d}h=dh\wedge\tilde{\ast}\tilde{d}f=[af_2h_2+cf_1h_1-b(f_1h_2+f_2h_1)]\omega_1\wedge\omega_2$
through a few steps of calculations. Therefore, We can arrive at
Lemma \ref{lemma1} by (\ref{midequ1}) minus (\ref{midequ2}).$\P$}

It follows that
\bequ\int_Mfd\tilde{\ast}\tilde{d}h=\int_Mhd\tilde{\ast}\tilde{d}f\label{greenident22}\eequ
for a closed surface.

Because $d\tilde{\ast}\tilde{d}f$ is a 2-form, we can define an
operator $\nabla\cdot\tilde{\nabla}$ such that
$d\tilde{\ast}\tilde{d}f=\nabla\cdot\tilde{\nabla}f\omega_1\wedge\omega_2$.

\subsection{Variational theory of surface}

If let $M$ undergoes an infinitesimal deformation such that every
point $\mathbf{r}$ of $M$ has a displacement $\delta\mathbf{r}$,
we obtain a new surface
$M'=\{\mathbf{r}'|\mathbf{r}'=\mathbf{r}+\delta\mathbf{r}\}$.
$\delta\mathbf{r}$ is called the variation of surface $M$ and
expressed as \beqn \delta
\mathbf{r}=\delta_1\mathbf{r}+\delta_2\mathbf{r}+\delta_3\mathbf{r},\label{deltar}\\
\delta_i\mathbf{r}=\Omega_{i}\mathbf{e}_{i}\quad
(i=1,2,3),\label{deltari} \eeqn where the repeated subindexes do
not represent Einstein summation.

{\defn If $f$ is a generalized function of $\mathbf{r}$
{\rm(}including scalar function, vector function, and $r$-form
dependent on point $\mathbf{r}${\rm)}, define \bequ\label{deltaif}
\delta_i^{(q)}
f=(q!)\mathcal{L}^{(q)}[f(\mathbf{r}+\delta_i\mathbf{r})-f(\mathbf{r})]\quad
(i=1,2,3;q=1,2,3,\cdots), \eequ and the $q$-order variation of $f$
\bequ\label{deltaf} \delta^{(q)}
f=(q!)\mathcal{L}^{(q)}[f(\mathbf{r}+\delta\mathbf{r})-f(\mathbf{r})]\quad(q=1,2,3,\cdots),
\eequ where $\mathcal{L}^{(q)}[\cdots]$ represents the terms of
$\Omega_1^{q_1}\Omega_2^{q_2}\Omega_3^{q_3}$ in Taylor series of
$[\cdots]$ with $q_1+q_2+q_3=q$ and $q_1,q_2,q_3$ being
non-negative integers.}

It is easy to prove that:

(i) \ $\delta_i^{(q)}$ and $\delta^{(q)}$ $(i=1,2,3;
q=1,2,\cdots)$ are linear operators;

(ii) \ $\delta_1^{(1)}$, $\delta_2^{(1)}$, $\delta_3^{(1)}$ and
$\delta^{(1)}$ are commutative with each other;

(iii) \ $\delta_i^{(q+1)}=\delta_i^{(1)}\delta_i^{(q)}$ and
$\delta^{(q+1)}=\delta^{(1)}\delta^{(q)}$, thus we can safely
replace $\delta_i^{(1)}$, $\delta_i^{(q)}$, $\delta^{(1)}$, and
$\delta^{(q)}$ by $\delta_i$, $\delta_i^q$, $\delta$, and
$\delta^q$ $(q=2,3,\cdots)$, respectively;

(iv) \ For functions $f$ and $g$, $\delta_i [f(\mathbf{r})\circ
g(\mathbf{r})] =\delta_i f(\mathbf{r})\circ
g(\mathbf{r})+f(\mathbf{r})\circ\delta_i g(\mathbf{r})$, where
$\circ$ represents the ordinary production, vector production or
exterior production;

(v) \ $\delta_i f[g(\mathbf{r})]=(\partial f/\partial g)\delta_i
g$;

(vi) $\delta^{q}=(\delta_1+\delta_2+\delta_3)^q$, e.g.
$\delta^{2}=\delta_1^2+\delta_2^2+\delta_3^2+2\delta_1\delta_2+2\delta_2\delta_3+2\delta_1\delta_3$.\\

Due to the deformation of $M$, the vectors
${\mathbf{e}_1,\mathbf{e}_2,\mathbf{e}_3}$ are also changed.
Denote their changes \bequ \delta_l \mathbf{e}_{i}=\Omega
_{lij}\mathbf{e}_{j},\eequ Obviously,
$\mathbf{e}_{i}\cdot\mathbf{e}_{j}=\delta_{ij}$ implies $\Omega
_{lij}=-\Omega _{lji}$. Because $\delta_1$, $\delta_2$ and
$\delta_3$ are linear mappings from $M$ to $M'$, they are
commutative with exterior differential operator $d$ \cite{chen2}.
Therefore, using $d\delta_l \mathbf{r}=\delta_l d\mathbf{r}$ and
$d\delta_l \mathbf{e}_j=\delta_l d\mathbf{e}_j$, we arrive at
\beqn
\delta_1 \omega _{1} =d\Omega _{1}-\omega _{2}\Omega _{121},\label{omvaratione11}\\
\delta_1 \omega _{2} =\Omega _{1}\omega _{12}-\omega _{1}\Omega _{112},\label{omvaratione12}\\
\Omega _{113}=a\Omega _{1},\quad\Omega _{123}=b\Omega
_{1};\label{omvaratione13} \eeqn
\beqn
\delta _{2}\omega _{1} =\Omega _{2}\omega _{21}-\omega _{2}\Omega _{221},\label{omvaratione21} \\
\delta _{2}\omega _{2} =d\Omega _{2}-\omega _{1}\Omega _{212},\label{omvaratione22} \\
\Omega _{213} =b\Omega _{2},\quad\Omega _{223}=c\Omega
_{2};\label{omvaratione23} \eeqn \beqn \delta_3 \omega _{1}
=\Omega _{3}\omega _{31}-\omega _{2}\Omega _{321},
\label{detaomega1} \\
\delta_3 \omega _{2} =\Omega _{3}\omega _{32}-\omega _{1}\Omega
_{312},
\label{detaomega2} \\
d\Omega _{3} =\Omega _{313}\omega _{1}+\Omega _{323}\omega _{2};
\label{domega3}\eeqn
 \beqn\delta_l \omega _{ij}=d\Omega _{lij}+\Omega _{lik}\omega
_{kj}-\omega _{ik}\Omega _{lkj}.\label{detaomegaij}\eeqn

Above equations (\ref{omvaratione11}) $\sim$ (\ref{detaomegaij})
are the fundamental equations in our paper and have not existed in
previous mathematical literature \cite{Griffiths} and
\cite{Bryant}. Otherwise, it is easy to deduce that
$\delta_i\tilde{d}f=\tilde{d}\delta_if$ $(i=1,2,3)$ for function
$f$.

\section{Closed lipid bilayers\label{close-bilayer}}
In this section, we will discuss the equilibrium shapes and
mechanical stabilities of closed lipid bilayers. We just consider
the closed surface in this section.
\subsection{First order variational problems on a closed surface}
In this subsection, we will discuss the first order variation of
the functional \bequ \mathcal{F}=\int_M
\mathcal{E}(2H[\mathbf{r}],K[\mathbf{r}])dA+p\int_V
dV,\label{funct1} \eequ where $H$ and $K$ are mean and gaussian
curvatures at point $\mathbf{r}$ in surface $M$. $p$ is a constant
and $V$ be the volume enclosed within the surface.

According to the variational theory of surface in
Sec.\ref{Math-pre}, we have
$\delta\mathcal{F}=\delta_1\mathcal{F}+\delta_2\mathcal{F}+\delta_3\mathcal{F}$.
Therefore, the next tasks are to calculate $\delta_1\mathcal{F}$,
$\delta_2\mathcal{F}$ and $\delta_3\mathcal{F}$, respectively.

\subsubsection{Calculation of $\delta_3\mathcal{F}$}\hspace{1cm}
Here, we will briefly prove 4 Lemmas and a theorem. Above all,
denote \bequ\mathcal{F}_{e}=\int_M
\mathcal{E}(2H[\mathbf{r}],K[\mathbf{r}])dA.\label{functfe}\eequ

{\lem $\delta_3 dA=-(2H)\Omega _{3}dA$.\label{delta3da}}\\
 {\prf $\delta_3 dA=\delta_3(\omega _{1}\wedge \omega
_{2})=\delta_3 \omega _{1}\wedge \omega _{2}+\omega _{1}\wedge
\delta_3 \omega _{2}$. Considering (\ref{omega13}), (\ref{meanh}),
(\ref{detaomega1}) and (\ref{detaomega2}), we arrive at this
Lemma. $\P$}

{\lem $\delta_3 (2H) dA=2(2H^{2}-K)\Omega _{3}dA+d\ast d\Omega_3$.\label{delta32H}}\\
{\prf $\delta_3 (2H) dA=\delta_3 a\omega _{1}\wedge
\omega_2+\delta_3 c\omega _{1}\wedge \omega_2$. Let $\delta_3$
acts on (\ref{omega13}), we have
\begin{eqnarray*}\delta_3
\omega _{13}&=&\delta_3 a\omega _{1}+a\delta_3 \omega
_{1}+\delta_3
b\omega _{2}+b\delta_3 \omega _{2},\\
\delta_3 \omega _{23}&=&\delta_3 b\omega _{1}+b\delta_3 \omega
_{1}+\delta_3 c\omega _{2}+c\delta_3 \omega _{2}.\end{eqnarray*}
If considering (\ref{meanh}), (\ref{gassianK}), (\ref{stardf}),
(\ref{detaomega1})$\sim$(\ref{detaomegaij}), we arrive at this
Lemma. $\P$}

{\lem $\delta_3 K
dA=2KH\Omega_3dA+d\tilde{\ast}\tilde{d}\Omega_{3}$.\label{delta3K}}\\
{\prf \emph{Theorem Egregium} (see \ref{appgbf}) implies that
$\delta_3 KdA=-\delta_3 d\omega_{12}-K\delta_3
dA=-d\delta_3\omega_{12}-K\delta_3 dA$. We will arrive at this
Lemma from (\ref{detaomegaij}) and Lemma \ref{delta3da} as well as
the discussions in Sec.\ref{gaussmapping}. $\P$}

{\thm
$\delta_3\mathcal{F}_e=\int_M[(\nabla^2+4H^2-2K)\frac{\partial\mathcal{E}}{\partial
(2H)}+(\nabla\cdot\tilde{\nabla}+2KH)\frac{\partial\mathcal{E}}{\partial
K}-2H\mathcal{E}]\Omega_3 dA$.\label{thedelta3fun}}\\
{\prf Above all, we have \beqn
\delta_3\mathcal{F}_e&=&\int_M\delta_3 \mathcal{E} dA+\int_M
\mathcal{E} \delta_3 A\nonumber\\&=&\int_M\frac{\partial
\mathcal{E}}{\partial (2H)}\delta_3 (2H) dA+\int_M\frac{\partial
\mathcal{E}}{\partial K}\delta_3 K dA+\int_M \mathcal{E}\delta_3
dA\nonumber. \eeqn

By using Lemmas \ref{delta3da}, \ref{delta32H}, \ref{delta3K}, we
obtain \beqn \delta_3\mathcal{F}_e&=&
\int_M\left[(4H^2-2K)\frac{\partial\mathcal{E}}{\partial
(2H)}+2KH\frac{\partial\mathcal{E}}{\partial
K}-2H\mathcal{E}\right]\Omega_3dA\nonumber\\
&+&\int_M \left[\frac{\partial\mathcal{E}}{\partial (2H)}d\ast
d\Omega_3+\frac{\partial \mathcal{E}}{\partial
K}d\tilde{\ast}\tilde{d}\Omega_{3}\right].\label{delta3func}\eeqn
For the closed surface $M$, we arrive at this theorem by
considering (\ref{stoke3}) and (\ref{greenident22}). $\P$}

{\lem $\delta_3\int_V dV=\int_M \Omega_{3}dA$.\label{delta3v}}\\
{\prf Because $M$ is a closed surface in $\mathbb{E}^3$, Stokes
theorem (see \ref{appdiff}) implies $\int_V 3 dV=\int_V
\nabla\cdot \mathbf{r}dV=\int_{\partial V}\mathbf{r}\cdot
\mathbf{n}dA$, thus \beqn
\delta_3 \int_V dV &=&\frac{1}{3}\int_M \delta_3 \lbrack \mathbf{r}\cdot \mathbf{e}%
_{3}(\omega _{1}\wedge \omega _{2})]\label{deltadv0} \\
&=&\frac{1}{3}\int_M [\delta_3 \mathbf{r}\cdot
\mathbf{e}_{3}(\omega _{1}\wedge \omega _{2})+\mathbf{r}\cdot
\delta_3 \mathbf{e}_{3}(\omega _{1}\wedge \omega
_{2})+\mathbf{r}\cdot \mathbf{e}_{3}\delta_3 (\omega _{1}\wedge
\omega _{2})]. \nonumber\eeqn

From (\ref{detaomega1})$\sim$(\ref{domega3}), we obtain
\beqn\delta_3 \mathbf{r}\cdot \mathbf{e}_{3}(\omega _{1}\wedge
\omega _{2})
&=&\Omega _{3}\omega _{1}\wedge \omega _{2},\label{deltadv1}\\
\mathbf{r}\cdot \mathbf{e}_{3}\delta_3 (\omega _{1}\wedge \omega
_{2})&=& \mathbf{r}\cdot \mathbf{e}_{3}(-2H)\Omega _{3}\omega
_{1}\wedge \omega _{2},\label{deltadv2}
\\
\mathbf{r}\cdot \delta_3 \mathbf{e}_{3}(\omega _{1}\wedge \omega
_{2})&=& -d\Omega _{3}\wedge (-\mathbf{r}\cdot \mathbf{e}
_{2}\omega _{1}+\mathbf{r}\cdot \mathbf{e}_{1}\omega
_{2}).\label{deltadv3}\eeqn By using the integration by parts and
Stokes theorem, we have \beqn -\int_M d\Omega _{3}\wedge
(-\mathbf{r}\cdot \mathbf{e} _{2}\omega _{1}+\mathbf{r}\cdot
\mathbf{e}_{1}\omega _{2}) &=&\int_M \Omega _{3}d(-\mathbf{r}\cdot
\mathbf{e} _{2}\omega _{1}+\mathbf{r}\cdot \mathbf{e}_{1}\omega
_{2})\nonumber\\&=&\int_M\Omega _{3}[2+\mathbf{r}\cdot
\mathbf{e}_{3}(2H)]\omega _{1}\wedge \omega _{2}.\label{deltadv4}
\eeqn

Therefore, we will arrive at $\delta_3\int_V dV=\int_M
\Omega_{3}dA$ by using (\ref{deltadv0})$\sim$(\ref{deltadv4}).
$\P$}

\subsubsection{Calculation of $\delta_1\mathcal{F}$ and $\delta_2\mathcal{F}$}
{\thm $\delta_1\mathcal{F}\equiv 0$ and $\delta_2\mathcal{F}\equiv
0$.\label{thedelta12fun}}\\
{\prf We obtain \bequ
db\wedge\omega_1+2bd\omega_1=(a-c)d\omega_2-dc\wedge\omega_2.\label{tempdomg23}
\eequ from (\ref{domgaij}) and (\ref{omega13}).

Using (\ref{omvaratione11})$\sim$(\ref{omvaratione13}),
(\ref{detaomegaij}), and (\ref{tempdomg23}), we arrive at \beqn
\delta_1 (\omega _{1}\wedge \omega _{2})=d(\Omega _{1}\omega _{2}),\label{detaae1}\\
\delta_1 (2H)\omega _{1}\wedge \omega _{2}=d(2H)\wedge \omega
_{2}\Omega _{1}\label{detaHe1} \eeqn through a few calculations.

By analogy with the proof of Lemma \ref{delta3K}, we can prove
that \bequ \delta_1 K\omega _{1}\wedge \omega _{2}=dK\wedge \Omega
_{1}\omega _{2}.\label{detaKe1} \eequ

Therefore, we have \beqn \delta_1
\mathcal{F}_e&=&\int_M\left[\frac{\partial \mathcal{E}}{\partial
(2H)}\delta_1 (2H)\omega _{1}\wedge \omega _{2}+\frac{\partial
\mathcal{E}}{\partial K}\delta_1 K\omega _{1}\wedge \omega
_{2}+\mathcal{E}\delta_1
(\omega _{1}\wedge \omega _{2})\right]\nonumber\\
&=&\int_M d(\mathcal{E}\omega _{2}\Omega _{1}).\label{delta1func}
\eeqn

Similarly, we can obtain \bequ \delta_2 \mathcal{F}_e=-\int_M
d(\mathcal{E}\omega _{1}\Omega _{2}).\label{delta2func}\eequ

Otherwise, it is not hard to obtain
$\delta_1\int_VdV=\int_Md(\textbf{r}\cdot\textbf{e}_3\omega
_{2}\Omega _{1})$ and
$\delta_2\int_VdV=-\int_Md(\textbf{r}\cdot\textbf{e}_3\omega
_{1}\Omega _{2})$.

Therefore, $\delta_1\mathcal{F}=\delta_2\mathcal{F}\equiv 0$
because $M$ is a closed surface. $\P$}

\subsubsection{Euler-Lagrange equation}\hspace{1cm}
Till now, we can obtain \bequ \delta\mathcal{F}
=\int_M\left[(\nabla^2+4H^2-2K)\frac{\partial\mathcal{E}}{\partial
(2H)}+(\nabla\cdot\tilde{\nabla}+2KH)\frac{\partial\mathcal{E}}{\partial
K}-2H\mathcal{E}+p\right]\Omega_3 dA.\label{deltafunc}\eequ Thus
the Euler-Lagrange equation corresponding to the functional
$\mathcal{F}$ is:
\begin{equation}\label{equalium}
\left[\left(\nabla^2+4H^2-2K\right)\frac{\partial}{\partial
(2H)}+\left(\nabla\cdot\tilde{\nabla}+2KH\right)\frac{\partial}{\partial
K}-2H\right]\mathcal{E}(2H,K)+p=0.\end{equation} The similar
equation is first found in Ref. \cite{natio}.

\subsection{Second order variation}
In this subsection, we discuss the second order variation of
functional (\ref{funct1}) . This problem was also studied by
Capovilla and Guven in recent paper \cite{Guvencap}. Because
$\delta_1\mathcal{F}=\delta_2\mathcal{F}\equiv 0$ for closed
surface $M$, we have
$\delta\delta_1\mathcal{F}=\delta\delta_2\mathcal{F}=0$, and
$\delta^2\mathcal{F}=\delta\delta_3\mathcal{F}=\delta_3^2\mathcal{F}$.

Form (\ref{delta3func}) and Lemma \ref{delta3v}, we obtain \beqn
\delta ^{2}\mathcal{F} &=&\delta _{3}\int_{M}\left[ (4H^{2}-2K)\frac{%
\partial \mathcal{E}}{\partial (2H)}+(2KH)\frac{\partial \mathcal{E}}{%
\partial K}-2H\mathcal{E}+p\right] \Omega _{3}dA\nonumber \\
&&+\delta _{3}\int_{M}\frac{\partial \mathcal{E}}{\partial
(2H)}d\ast
d\Omega _{3}+\delta _{3}\int_{M}\frac{\partial \mathcal{E}}{\partial K}d%
\tilde{\ast}\tilde{d}\Omega _{3} \nonumber\\
&=&\int_{M}\delta _{3}\left[ (4H^{2}-2K)\frac{\partial
\mathcal{E}}{\partial (2H)}+(2KH)\frac{\partial
\mathcal{E}}{\partial K}-2H\mathcal{E}+p\right]
\Omega _{3}dA\nonumber \\
&&+\int_{M}\left[ (4H^{2}-2K)\frac{\partial \mathcal{E}}{\partial (2H)}+(2KH)%
\frac{\partial \mathcal{E}}{\partial K}-2H\mathcal{E}+p\right]
\Omega
_{3}\delta _{3}dA \nonumber\\
&&+\int_{M}\delta _{3}\left[ \frac{\partial \mathcal{E}}{\partial (2H)}%
\right] d\ast d\Omega _{3}+\frac{\partial \mathcal{E}}{\partial
(2H)}\delta
_{3}(d\ast d\Omega _{3})\nonumber \\
&&+\int_{M}\delta _{3}\left( \frac{\partial \mathcal{E}}{\partial K}\right) d%
\tilde{\ast}\tilde{d}\Omega _{3}+\frac{\partial \mathcal{E}}{%
\partial K}\delta _{3}(d\tilde{\ast}\tilde{d}\Omega _{3}).\label{secondvariation}
\eeqn Please notice that $\Omega _{3}$ can freely come into and
out of the expressions acted by the operator $\delta_3$.

{\lem For every function $f$, $\delta _{3}d\ast df=d\ast d\delta
_{3}f+d(2H\Omega _{3}\ast df)-2d(\Omega
_{3}\tilde{\ast}\tilde{d}f)$.\label{delta3dsdf}}\\
{\prf Let $df =f_{1}\omega _{1}+f_{2}\omega _{2}$, we have $\ast
df =f_{1}\omega _{2}-f_{2}\omega _{1}$, $\tilde{d}f =f_{1}\omega
_{13}+f_{2}\omega _{23}$ and $\tilde{\ast} \tilde{d}f =f_{1}\omega
_{23}-f_{2}\omega _{13}$. By using (\ref{detaomega1}) and
(\ref{detaomega2}), we have \beqn \delta _{3}\ast df=(\delta
_{3}f_{1}\omega _{2}-\delta _{3}f_{2}\omega _{1})-\Omega
_{312}df+\Omega _{3}[f_{2}(a\omega _{1}+b\omega
_{2})-f_{1}(b\omega _{1}+c\omega _{2})],\nonumber\\
\ast \delta _{3}df =(\delta _{3}f_{1}\omega _{2}-\delta
_{3}f_{2}\omega _{1})-\Omega _{312}df+\Omega _{3}[f_{1}(b\omega
_{1}-a\omega _{2})+f_{2}(c\omega _{1}-b\omega
_{2})]\nonumber,\\
\delta _{3}\ast df-\ast \delta _{3}df =2H\Omega _{3}\ast df-2\Omega _{3}%
\tilde{\ast}\tilde{d}f.\label{delta3dsdftemp}\eeqn Using the
operator $d$ to act on both sides of (\ref{delta3dsdftemp}) and
noticing the commutativity of $d$ and $\delta_3$, we arrive at
this Lemma. \P}

{\lem For every function $f$, $\delta
_{3}d\tilde{\ast}\tilde{d}f=d[\delta _{3}(2H)\ast df+2H\delta
_{3}\ast df+2K\Omega _{3}\ast df-2H\Omega _{3}\ast \tilde{d}f-\ast
\tilde{d}\delta _{3}f]$.\label{delta3dtsdf}}\\
{\prf Similar to the proof of Lemma \ref{delta3dsdf}, we have
\beqn \delta _{3}\ast \tilde{d}f=\delta _{3}(af_{1}+bf_{2})\omega
_{2}-\delta _{3}(bf_{1}+cf_{2})\omega _{1}-K\Omega _{3}\ast
df-\Omega
_{312}\tilde{d}f, \nonumber\\
\ast \delta _{3}\tilde{d}f=\delta _{3}(af_{1}+bf_{2})\omega
_{2}-\delta _{3}(bf_{1}+cf_{2})\omega _{1}-2H\Omega _{3}\ast
\tilde{d}f+K\Omega _{3}\ast df-\Omega _{312}\tilde{d}f.
\nonumber\eeqn The difference of above two equations gives
\bequ\ast \delta _{3}\tilde{d}f-\delta _{3}\ast
\tilde{d}f=2K\Omega _{3}\ast df-2H\Omega _{3}\ast
\tilde{d}f.\label{delta3dtsdftemp1}\eequ Otherwise, It is easy to
see \bequ \ast \tilde{d}f+\tilde{\ast}\tilde{d}f =2H\ast
df.\label{delta3dtsdftemp2}\eequ Using $d\delta_3$ to act on both
sides of (\ref{delta3dtsdftemp2}) and $d$ to act on both sides of
(\ref{delta3dtsdftemp1}), considering the commutative relations
$\tilde{d}\delta_3=\delta_3\tilde{d}$ and $d\delta_3=\delta_3d$,
we arrive at this Lemma. \P}

If $df=f_1\omega_1+f_2\omega_2$, we define $\nabla
f=f_1\mathbf{e}_1+f_2\mathbf{e}_2$, $\bar{\nabla}f
=(af_{1}+bf_{2})\mathbf{e}_{1}+(bf_{1}+cf_{2})\mathbf{e}_{2}$,
$\tilde{\nabla}f
=(cf_{1}-bf_{2})\mathbf{e}_{1}+(af_{2}-bf_{1})\mathbf{e}_{2}$ and
$d\ast
\tilde{d}f=(\nabla\cdot\bar{\nabla})f\omega_1\wedge\omega_2$. It
follows that, for function $f$ and $g$, \beqn
\tilde{\nabla}f+\bar{\nabla}f=2H\nabla
f\label{nablarelation}\\df\wedge \ast dg=(\nabla f\cdot\nabla
g)\omega_1\wedge\omega_2,\label{nabladnabla}\\
df\wedge \ast \tilde{d}g=(\nabla f\cdot\bar{\nabla}
g)\omega_1\wedge\omega_2,\label{nablabnabla}\\
df\wedge \tilde{\ast} \tilde{d}g=(\nabla f\cdot\tilde{\nabla}
g)\omega_1\wedge\omega_2.\label{nablatnabla}
 \eeqn

{\remk The tensor expressions of $\nabla$, $\bar{\nabla}$,
$\tilde{\nabla}$, $\nabla^2$, $\nabla\cdot\bar{\nabla}$,
$\nabla\cdot\tilde{\nabla}$ are developed in {\rm\ref{app2}}.}

 {\thm The second order variation of functional
{\rm(\ref{funct1})} is \beqn \delta^2 \mathcal{F}&=&\int_M\Omega
_{3}^{2}\left[(4H^{2}-2K)^{2}\frac{\partial
^{2}\mathcal{E}}{\partial (2H)^{2}} -4KH\frac{\partial
\mathcal{E}}{\partial (2H)}-2K^{2}%
\frac{\partial \mathcal{E}}{\partial K}\right.\nonumber\\&&\quad+\left.4KH(4H^{2}-2K)\frac{\partial ^{2}%
\mathcal{E}}{\partial (2H)\partial K}+4K^{2}H^{2}\frac{\partial ^{2}\mathcal{%
E}}{\partial K^{2}}+2K\mathcal{E}-2Hp\right]dA \nonumber\\
&+&\int_M\Omega _{3}\nabla ^{2}\Omega _{3}\left[4H\frac{\partial
\mathcal{E}}{\partial
(2H)}+4(2H^{2}-K)\frac{\partial ^{2}\mathcal{E}}{\partial (2H)^{2}}+K\frac{%
\partial \mathcal{E}}{\partial K}\right.\nonumber\\&&\quad+\left.4HK\frac{\partial ^{2}\mathcal{E}}{%
\partial K\partial (2H)}-\mathcal{E}+8H^{2}\frac{\partial \mathcal{E}}{%
\partial K}\right]dA \nonumber\\
&+&\int_M\Omega _{3}\nabla \cdot \tilde{\nabla}\Omega _{3}\left[4(2H^{2}-K)\frac{%
\partial ^{2}\mathcal{E}}{\partial (2H)\partial K}-4\frac{\partial \mathcal{E%
}}{\partial (2H)}\right.\nonumber\\&&\quad+\left.4HK\frac{\partial ^{2}\mathcal{E}}{\partial K^{2}}-4H\frac{%
\partial \mathcal{E}}{\partial K}\right]dA \nonumber\\
&+&\int_M(\nabla ^{2}\Omega _{3})^{2}\left[\frac{\partial
^{2}\mathcal{E}}{\partial
(2H)^{2}}+\frac{\partial \mathcal{E}}{\partial K}\right]dA\nonumber\\
&+&\int_M\left[\frac{2\partial ^{2}%
\mathcal{E}}{\partial (2H)\partial K}\nabla ^{2}\Omega _{3}\nabla
\cdot \tilde{\nabla}\Omega _{3}+\frac{\partial
\mathcal{E}}{\partial (2H)}\nabla
(2H\Omega _{3})\cdot \nabla \Omega _{3}\right]dA \nonumber\\
&+&\int_M\left[\frac{\partial ^{2}\mathcal{E}}{\partial
K^{2}}(\nabla
\cdot \tilde{\nabla}\Omega _{3})^{2}-\frac{2\partial \mathcal{E}}{\partial (2H)}\nabla \Omega _{3}\cdot \tilde{%
\nabla}\Omega _{3}dA\right]\nonumber\\&+&\int_M\frac{\partial \mathcal{E}}{\partial K%
}\left[\nabla (8H^{2}\Omega _{3}+\nabla ^{2}\Omega _{3})\cdot
\nabla \Omega _{3}-\nabla (4H\Omega _{3})\cdot
\tilde{\nabla}\Omega _{3}-4H\Omega _{3}\nabla \cdot
\tilde{\nabla}\Omega _{3}\right.\nonumber\\
&&\quad-\left.\nabla (2H\Omega _{3})\cdot \bar{\nabla}\Omega
_{3}-2H\Omega _{3}\nabla \cdot \bar{\nabla}\Omega
_{3}\right]dA.\nonumber\eeqn\label{thvar2Fc}}\\
 {\prf Replacing $f$ by $\Omega_3$ in
Lemma \ref{delta3dsdf} and \ref{delta3dtsdf}, and noticing that
$\Omega_3$ is similar to a constant relative to $\delta_3$, we
have \beqn &&\delta _{3}d\ast d\Omega _{3}=[\nabla (2H\Omega
_{3})\cdot \nabla \Omega _{3}+2H\Omega _{3}\nabla ^{2}\Omega
_{3}-2\nabla \Omega _{3}\cdot \tilde{\nabla}\Omega _{3}-2\Omega
_{3}\nabla \cdot \tilde{\nabla}\Omega _{3}]dA,\nonumber\\ &&\delta
_{3}d\tilde{\ast}\tilde{d}\Omega _{3} =[\nabla (8H^{2}\Omega
_{3}+\nabla ^{2}\Omega _{3})\cdot \nabla \Omega
_{3}+(8H^{2}\Omega _{3}+\nabla ^{2}\Omega _{3})\nabla ^{2}\Omega _{3}\nonumber \\
&&\quad-\nabla (4H\Omega _{3})\cdot \tilde{\nabla}\Omega
_{3}-4H\Omega _{3}\nabla
\cdot \tilde{\nabla}\Omega _{3}-\nabla (2H\Omega _{3})\cdot \bar{\nabla}%
\Omega _{3}-2H\Omega _{3}\nabla \cdot \bar{\nabla}\Omega
_{3}]dA.\nonumber \eeqn Substituting them into
(\ref{secondvariation}) and using Lemmas \ref{delta32H} and
\ref{delta3K}, we can arrive at this theorem through expatiatory
calculations. \P}

In particular, if $\partial \mathcal{E}/\partial K=\bar{k}$ being
a constant, (\ref{secondvariation}) is simplified to \beqn
\delta^2 \mathcal{F}&=&\int_M\Omega
_{3}^{2}\left[(4H^{2}-2K)^{2}\frac{\partial
^{2}\mathcal{E}_H}{\partial
(2H)^{2}}-4HK\frac{\partial \mathcal{E}_H}{\partial (2H)}+2K\mathcal{E}_H-2Hp\right]dA\nonumber \\
&&+\int_M\Omega _{3}\nabla ^{2}\Omega _{3}\left[4H\frac{\partial
\mathcal{E}_H}{\partial
(2H)}+4(2H^{2}-K)\frac{\partial ^{2}\mathcal{E}_H}{\partial (2H)^{2}}-\mathcal{%
E}_H\right]dA \nonumber\\
&&-\int_M\frac{4\partial \mathcal{E}_H}{\partial (2H)}\Omega _{3}\nabla \cdot \tilde{%
\nabla}\Omega _{3}dA+\int_M\frac{\partial ^{2}\mathcal{E}_H}{\partial (2H)^{2}}%
(\nabla ^{2}\Omega _{3})^{2}dA \nonumber\\
&&+\int_M\frac{\partial \mathcal{E}_H}{\partial (2H)}\left[\nabla
(2H\Omega _{3})\cdot \nabla \Omega _{3}-2\nabla \Omega _{3}\cdot
\tilde{\nabla}\Omega _{3}\right]dA,\label{secondvarnoK}\eeqn where
$\mathcal{E}_H=\mathcal{E}-\bar{k}K$.

\subsection{Shape equation of closed lipid bilayers}
Now, Let us turn to the shape equation of closed lipid bilayers.
We take the free energy of closed lipid bilayer under the osmotic
pressure as (\ref{free-e-closed}). Substituting
$\mathcal{E}=(k_c/2)(2H+c_0)^2+\bar{k}K+\mu$ into
(\ref{equalium}), we obtain the shape equation
(\ref{shape-closed}). This equation is the fourth order nonlinear
equation. It is not easy to obtain its special solutions. We will
give three typical analytical solutions as follows. Some new
important results on it can be found in recent paper by Landolfi
\cite{GLandolfi}.

\subsubsection{Constant mean curvature surface}\hspace{1cm}
From 1956 to 1958, Alexandrov proved an unexpected theorem: an
embedded surface (i.e. the surface does not intersect with itself)
with constant mean curvature in $\mathbb{E}^3$ must be a spherical
surface \cite{Alexandrov}. Thus the closed bilayer with constant
mean curvature must be a sphere. For a sphere with radius $R$, we
have $H=-1/R$ and $K=1/R^2$. Substituting them into
(\ref{shape-closed}), we arrive at \bequ pR^2+2\mu
R-k_cc_0(2-c_0R)=0.\label{sphericalbilayer}\eequ This equation
gives the spherical radius under the osmotic pressure $p$.

\subsubsection{Biconcave discoid shape and $\sqrt{2}$ torus}\hspace{1cm}
It is instructive to find some axisymmetrical solutions to the
shape equation (\ref{shape-closed}). To do that, we denote
$\mathbf{r}=\{u\cos v, u\sin v, z\}$,
$\psi=\arctan[\frac{dz(u)}{du}]$, and $\Psi=\sin\psi$. Thus
(\ref{shape-closed}) is transformed into \beqn
&&(\Psi ^{2}-1)\frac{d^{3}\Psi }{du^{3}}+\Psi \frac{d^{2}\Psi }{du^{2}}\frac{%
d\Psi }{du}-\frac{1}{2}\left(\frac{d\Psi }{du}\right)^{3}-\frac{p}{k_c}\nonumber\\&+&\frac{2(\Psi ^{2}-1)}{u}\frac{%
d^{2}\Psi }{du^{2}}+\frac{3\Psi }{2u}\left(\frac{d\Psi }{du}\right)^{2}+\left(\frac{c_{0}^{2}}{2}+\frac{2c_{0}\Psi }{u}+\frac{\mu }{k_{c}}-\frac{%
3\Psi ^{2}-2}{2u^{2}}\right)\frac{d\Psi }{du}\nonumber \\
&+&\left(\frac{c_{0}^{2}}{2}+\frac{\mu}{%
k_{c}}-\frac{1}{u^{2}}\right)\frac{\Psi }{u}+\frac{\Psi
^{3}}{2u^{3}}=0.\label{nequilbcl} \eeqn

To find the solution of (\ref{nequilbcl}) that satisfies $\Psi=0$
when $u=0$, we consider the asymptotic form of (\ref{nequilbcl})
at $u=0$: \bequ \frac{d^{3}\Psi
}{du^{3}}+\frac{2}{u}\frac{d^{2}\Psi
}{du^{2}}-\frac{1}{u^2}\frac{d\Psi }{du}+\frac{\Psi}{u^3}=0.\eequ
Please notice that there are two misprints in our previous paper
\cite{tzc1}. Above equation is the Euler differential equation and
has the general solution $\Psi=\alpha_1/u+\alpha_2u+\alpha_3u\ln
u$ with three integral constants $\alpha_1=0$, $\alpha_2$, and
$\alpha_3$. The asymptotic solution hints that
$\Psi=-c_0u\ln(u/u_B)$ might be a solution to (\ref{nequilbcl})
which requires $p=\mu=0$. When $0<c_0u_B<e$,
$\Psi=-c_0u\ln(u/u_B)$ corresponds to the biconcave discoid shape
\cite{oy1,oy3}.

Otherwise, when $\mu/k_c=-2\alpha c_{0}-c_{0}^{2}/2$ and
$p/k_{c}=-2\alpha^{2}c_{0}$, $\Psi=\alpha u+\sqrt{2}$ satisfies
(\ref{nequilbcl}). This solution corresponds to a torus with the
ratio of its two radii being exactly $\sqrt{2}$ if $\alpha<0$
\cite{oy1,oy4}.

\subsection{Mechanical stability of spherical bilayers}
A spherical bilayer can be described by
$\mathbf{r}=R(\sin\theta\cos\phi,sin\theta\sin\phi,cos\theta)$
with $R$ satisfying (\ref{sphericalbilayer}). We have $H=-1/R$,
$K=1/R^2$, $\tilde{\nabla}=-(1/R)\nabla$, $\nabla \cdot
\tilde{\nabla}=-(1/R)\nabla^2$ and
$\nabla^2=\frac{1}{R^2\sin\theta}\frac{\partial}{\partial\theta}
\left(\sin\theta\frac{\partial}{\partial\theta}\right)
+\frac{1}{R^2\sin^2\theta}\frac{\partial^2}{\partial\phi^2}$. If
we take $\mathcal{E}_H=(k_c/2)(2H+c_0)^2+\mu$,
(\ref{secondvarnoK}) is transformed into \beqn \delta
^{2}\mathcal{F}&=&(2c_{0}k_{c}/R+pR)\int_{0}^{\pi }\sin \theta
d\theta \int_{0}^{2\pi }d\phi \Omega
_{3}^{2}\nonumber\\&+&(k_{c}c_{0}R+2k_{c}+pR^{3}/2)\int_{0}^{\pi
}\sin \theta d\theta \int_{0}^{2\pi }d\phi\Omega _{3}\nabla
^{2}\Omega _{3}\nonumber\\&+&k_{c}R^{2}\int_{0}^{\pi }\sin \theta
d\theta \int_{0}^{2\pi }d\phi (\nabla ^{2}\Omega
_{3})^{2}.\label{secondvsph}\eeqn Expand $\Omega _{3}$ with the
spherical harmonic functions \cite{wangzx}: \bequ\Omega _{3}
=\sum_{l=0}^{\infty
}\sum_{m=-l}^{m=l}a_{lm}Y_{lm}(\theta,\phi),\quad a_{lm}^{*}
=(-1)^{m}a_{l,-m}.\label{harmonicf}\eequ If considering $\nabla
^{2}Y_{lm} =-l(l+1)Y_{lm}/R^{2}$ and $\int_{0}^{\pi}\sin \theta
d\theta\int_{0}^{2\pi}d\phi Y_{lm}^{*}Y_{l^{\prime}m^{\prime}}
=\delta _{mm^{\prime }}\delta _{ll^{\prime}}$, we transform
(\ref{secondvsph}) into \bequ\delta^{2}\mathcal{F}=
(R/2)\sum_{l,m}|a_{lm}|^{2}[l(l+1)-2]\{2k_{c}/R^{3}[l(l+1)-c_{0}R]-p\}.\eequ
Denote that \bequ p_l=(2k_{c}/R^{3})[l(l+1)-c_{0}R]\quad
(l=2,3,\cdots).\label{criticalps}\eequ When $p>p_l$,
$\delta^{2}\mathcal{F}$ can take negative value. Therefore, we can
take the critical pressure as \bequ
p_c=\min\{p_l\}=p_2=(2k_{c}/R^{3})(6-c_{0}R).\label{criticalpsc}\eequ
In this case, the spherical bilayer will be inclined to transform
into the biconcave discoid shape.

\section{Open lipid bilayers\label{open-bilayer}}
In this section, we will deal with the variational problems on
surface $M$ with edge $C$ as shown in Fig.\ref{surfacem}, and
discuss the shape equation and boundary conditions of open lipid
bilayers with free edges.
\subsection{First order variational problems on an open surface}
In this subsection, we will discuss the first order variation of
the functional \bequ \mathcal{F}=\int_M
\mathcal{E}(2H[\mathbf{r}],K[\mathbf{r}])dA+\int_C\Gamma(k_n,k_g)ds.\label{functopen}
\eequ Denote $\mathcal{F}_e=\int_M
\mathcal{E}(2H[\mathbf{r}],K[\mathbf{r}])dA$, and
$\mathcal{F}_C=\int_C\Gamma(k_n,k_g)ds$.

In terms of \ref{appcurve}, we have $\omega_2=0$, $ds=\omega_1$,
$k_n=a$, $k_gds=\omega_{12}$, $\tau_g=b$ in curve $C$. Using
(\ref{omvaratione11})$\sim$(\ref{detaomegaij}), we can arrive at
\beqn \delta _{1}\mathcal{F}_{C}=\int_{C}d(\Gamma \Omega
_{1})=0,\label{delta1funcopenc}\eeqn \beqn
\delta _{2}\mathcal{F}_{C}&=&\int_{C}\left[ \frac{d^{2}}{ds^{2}}\left( \frac{%
\partial \Gamma }{\partial k_{g}}\right) +K\frac{\partial \Gamma }{\partial
k_{g}}-k_{g}\left( \Gamma -\frac{\partial \Gamma }{\partial k_{g}}%
k_{g}\right)\right.\nonumber\\ &&\quad+\left.2(k_{n}-H)k_{g}\frac{\partial \Gamma }{\partial k_{n}}-\tau_g\frac{%
d}{ds}\left( \frac{\partial \Gamma }{\partial k_{n}}\right) -\frac{d}{ds}%
\left(\tau_g\frac{\partial \Gamma }{\partial k_{n}}\right) \right]
\Omega _{2}ds, \label{delta2funcopenc}\eeqn \beqn\delta
_{3}\mathcal{F}_{C} &=&\int_{C}\left[ \frac{d^{2}}{ds^{2}}\left(
\frac{\partial \Gamma }{\partial k_{n}}\right) +\frac{\partial \Gamma }{%
\partial k_{n}}(k_{n}^{2}-\tau _{g}^{2})+\tau _{g}\frac{d}{ds}\left( \frac{%
\partial \Gamma }{\partial k_{g}}\right)\right. \nonumber\\&&\quad+\left.\frac{d}{ds}\left( \tau _{g}\frac{%
\partial \Gamma }{\partial k_{g}}\right) -\left( \Gamma -\frac{\partial
\Gamma }{\partial k_{g}}k_{g}\right) k_{n}\right] \Omega _{3}ds\nonumber \\
&&\quad+\int_{C}\left( \frac{\partial \Gamma }{\partial k_{g}}k_{n}-\frac{%
\partial \Gamma }{\partial k_{n}}k_{g}\right) \Omega _{323}ds. \label{delta3funcopenc}\eeqn

In particular, (\ref{delta3func}), (\ref{delta1func}) and
(\ref{delta2func}) are still applicable. Consequently, \bequ\delta
_{1}\mathcal{F}_{e}=\int_{M}d(\mathcal{E}\omega _{2}\Omega
_{1})=\int_{C}\mathcal{E}\omega _{2}\Omega _{1}=0,\eequ \bequ
\delta _{2}\mathcal{F}_{e}=-\int_{M}d(\mathcal{E}\omega _{1}\Omega
_{2})=-\int_{C}\mathcal{E}\Omega _{2}ds, \eequ \beqn\delta _{3}\mathcal{F}_{e} &=&\int_{M}\left[ (\nabla ^{2}+4H^{2}-2K)\frac{%
\partial \mathcal{E}}{\partial (2H)}+(\nabla \cdot \tilde{\nabla}+2KH)\frac{%
\partial \mathcal{E}}{\partial K}-2H\mathcal{E}\right] \Omega _{3}dA \nonumber\\
&&+\int_{C}\left[ \mathbf{e}_{2}\cdot \nabla \left[ \frac{\partial \mathcal{E%
}}{\partial (2H)}\right] +\mathbf{e}_{2}\cdot \tilde{\nabla}\left( \frac{%
\partial \mathcal{E}}{\partial K}\right) -\frac{d}{ds}\left( \frac{\partial
\mathcal{E}}{\partial K}\right) \right] \Omega _{3}ds \nonumber\\
&&+\int_{C}\left[-\frac{\partial \mathcal{E}}{\partial (2H)}-k_{n}\frac{%
\partial \mathcal{E}}{\partial K}\right] \Omega _{323}ds.\eeqn

The functions $\Omega _{323}$, $\Omega _{2}$, and $\Omega _{3}$
can be regarded as virtual displacements. Thus $\delta
\mathcal{F}=(\delta_1+\delta_2+\delta_3)(\mathcal{F}_e+\mathcal{F}_C)=0$
gives
\begin{eqnarray}
&&(\nabla ^{2}+4H^{2}-2K)\frac{\partial \mathcal{E}}{\partial
(2H)}+(\nabla
\cdot \tilde{\nabla}+2KH)\frac{\partial \mathcal{E}}{\partial K}-2H\mathcal{E%
}=0,\label{euleropen1} \\
&&\mathbf{e}_{2}\cdot \nabla \left[ \frac{\partial \mathcal{E}}{\partial (2H)%
}\right] +\mathbf{e}_{2}\cdot \tilde{\nabla}\left( \frac{\partial \mathcal{E}%
}{\partial K}\right) -\frac{d}{ds}\left(\tau_g \frac{\partial \mathcal{E}}{%
\partial K}\right) +\frac{d^{2}}{ds^{2}}\left( \frac{\partial \Gamma }{%
\partial k_{n}}\right) +\frac{\partial \Gamma }{\partial k_{n}}%
(k_{n}^{2}-\tau _{g}^{2})\nonumber \\
&&\quad +\tau _{g}\frac{d}{ds}\left( \frac{\partial \Gamma }{\partial k_{g}}%
\right) +\frac{d}{ds}\left( \tau _{g}\frac{\partial \Gamma }{\partial k_{g}}%
\right) -\left. \left( \Gamma -\frac{\partial \Gamma }{\partial k_{g}}%
k_{g}\right) k_{n}\right\vert _{C}=0,\label{euleropen2} \\
&&-\frac{\partial \mathcal{E}}{\partial (2H)}-k_{n}\frac{\partial \mathcal{E}%
}{\partial K}+\frac{\partial \Gamma }{\partial k_{g}}k_{n}-\left. \frac{%
\partial \Gamma }{\partial k_{n}}k_{g}\right\vert _{C}=0,\label{euleropen3} \\
&&\frac{d^{2}}{ds^{2}}\left( \frac{\partial \Gamma }{\partial
k_{g}}\right) +K\frac{\partial \Gamma }{\partial
k_{g}}-k_{g}\left( \Gamma -\frac{\partial \Gamma }{\partial
k_{g}}k_{g}\right) +2(k_{n}-H)k_{g}\frac{\partial \Gamma
}{\partial k_{n}}\nonumber \\
&&\quad -\tau_g\frac{d}{ds}\left( \frac{\partial \Gamma }{\partial
k_{n}}\right)
\left. -\frac{d}{ds}\left( \tau_g\frac{\partial \Gamma }{\partial k_{n}}\right) -%
\mathcal{E}\right\vert _{C}=0.\label{euleropen4}
\end{eqnarray}
Among above equations, (\ref{euleropen1}) determines the shape of
the surface $M$, and (\ref{euleropen2})$\sim$(\ref{euleropen4})
determine the position of curve $C$ in the surface $M$.

\subsection{Shape equation and boundary conditions of open lipid bilayers}
In order to obtain the shape equation and boundary conditions of
an open lipid bilayer with an edge $C$, we take
$\mathcal{E}=(k_c/2)(2H+c_0)^2+\bar{k}K+\mu$ and $\Gamma
=\frac{1}{2}k_{b}(k_{n}^{2}+k_{g}^{2})+\gamma$ with $k_b$ and
$\gamma$ being constants. In this case,
(\ref{euleropen1})$\sim$(\ref{euleropen4}) are transformed into
\beqn
&& k_{c}(2H+c_{0})(2H^{2}-c_{0}H-2K)+k_{c}\nabla ^{2}(2H)-2\mu H =0,\label{shape-open}\\
&&k_{b}[d^{2}k_{n}/ds^{2}+k_{n}(\kappa ^{2}/2-\tau _{g}^{2})+\tau
_{g}dk_{g}/ds+d(\tau _{g}k_{g})/ds]\nonumber \\
&&\qquad+k_{c}\mathbf{e}_{2}\cdot \nabla (2H)-\left. \bar{k}d\tau
_{g}/ds-\gamma
k_{n}\right\vert _{C} =0, \label{boundcond1}\\
&&\left. k_{c}(2H+c_{0})+\bar{k}k_{n}\right\vert _{C} =0,\label{boundcond2} \\
&&k_{b}[d^{2}k_{g}/ds^{2}+k_{g}(\kappa ^{2}/2-\tau _{g}^{2})-\tau
_{g}dk_{n}/ds-d(\tau _{g}k_{n})/ds]\nonumber \\
&&\qquad-\left. [(k_{c}/2)(2H+c_{0})^{2}+\bar{k}K+\mu +\gamma
k_{g}]\right\vert _{C} =0,\label{boundcond3}\eeqn where
$\kappa^2=k_n^2+k_g^2$.

In fact, the above four equations express the force and moment
equilibrium equations of the surface and the edge:
(\ref{shape-open}) represents the force equilibrium equation of
point in the surface $M$ along $\mathbf{e}_3$ direction;
(\ref{boundcond1}) represents the force equilibrium equation of
point in the curve $C$ along $\mathbf{e}_3$ direction;
(\ref{boundcond2}) represents the bending moment equilibrium
equation of point in the curve $C$ around $\mathbf{e}_1$
direction; (\ref{boundcond3}) represents the force equilibrium
equation of point in the curve $C$ along $\mathbf{e}_2$ direction.

If $k_b=0$, (\ref{shape-open}) and (\ref{boundcond2}) remain
unchanged, but (\ref{boundcond1}) and (\ref{boundcond3}) are
simplified to \beqn \left. k_{c}\mathbf{e}_{2}\cdot \nabla
(2H)-\bar{k}d\tau _{g}/ds-\gamma
k_{n}\right\vert _{C} =0,\label{boundcond11} \\
\left. (k_{c}/2)(2H+c_{0})^{2}+\bar{k}K+\mu +\gamma
k_{g}\right\vert _{C} =0.\label{boundcond33} \eeqn

\subsection{Two-component lipid bilayer}
In this subsection, we study a closed bilayer consists of two
domains containing different kinds of lipid. This problem in
axisymmetrical case was theoretically discussed by J\"{u}licher
and Lipowsky \cite{Julicher}. The shapes of two-component bilayers
also were observed in recent experiment\cite{Baumgart}.

We assume that the boundary between two domains is a smooth curve
and the bilayer is still a smooth surface. The free energy is
written as \beqn \mathcal{F}=
p\int_VdV+\int_{M_I}[(k_c^I/2)(2H+c_0^I)^2+\bar{k}^IK+\mu^I]dA\nonumber\\\qquad+
\int_{M_{II}}[(k_c^{II}/2)(2H+c_0^{II})^2+\bar{k}^{II}K+\mu^{II}]dA
+\gamma\int_C ds.\label{freetwocom}\eeqn

In terms of the discussions on closed bilayers in section
\ref{close-bilayer} and above discussions in this section, we can
promptly write the shape equations of the two-component bilayer
without any symmetrical assumption as \bequ p-2\mu^i
H+k_c^i(2H+c_0)(2H^2-c_0^iH-2K)+k_c^i\nabla^2(2H)=0,
\label{shape-twocom}\eequ where the superscripts $i=I$ and $II$
represent the two lipid domains, respectively. And the boundary
conditions are as follows: \beqn \left. k_{c}^I\mathbf{e}_{2}\cdot
\nabla (2H)-\bar{k}^Id\tau _{g}/ds+k_{c}^{II}\mathbf{e}_{2}\cdot
\nabla (2H)-\bar{k}^{II}d\tau _{g}/ds-\gamma
k_{n}\right\vert _{C} =0,\label{boundtwocom1} \\
\left. k_{c}^I(2H+c_{0}^I)+\bar{k}^Ik_{n}
-[k_{c}^{II}(2H+c_{0}^{II})+\bar{k}^{II}k_{n}]\right\vert _{C}=0,\label{boundtwocom2} \\
(k_{c}^I/2)(2H+c_{0}^I)^{2}+\bar{k}^IK+\mu^I\nonumber\\
\qquad-\left.[(k_{c}^{II}/2)(2H+c_{0}^{II})^{2}+\bar{k}^{II}K+\mu^{II}]
+\gamma k_{g}\right\vert _{C} =0.\label{boundtwocom3} \eeqn

In above equations, the positive direction of curve $C$ is set to
along $\mathbf{e}_1$ of the lipid domain consisting of component
$I$. Furthermore, (\ref{shape-twocom})$\sim$(\ref{boundtwocom3})
are also applied to describe the closed bilayer with more than two
domains. But the boundary conditions are not applied to the
bilayer with a sharp angle across the boundary between domains.

\section{Cell membranes with cross-linking structures\label{cell-membrane}}
Cell membranes contain cross-linking protein structures. As is
well known, rubber also consists of cross-linking polymer
structures \cite{Treloar}. In this section, we first drive the
free energy of cell membrane with cross-linking protein structure
by analogy with the rubber elasticity. Secondly, we derive the
shape equation and in-plane stain equations of cell membrane by
taking the first order variation of the free energy. Lastly, we
discuss the mechanical stability of spherical cell membrane.

\subsection{The free energy of cell membrane}
Above all, we discuss the free energy change of a Gaussian chain
in a small strain field \bequ \epsilon =\left(
\begin{array}{ccc}
\varepsilon _{xx} & \varepsilon _{xy} & 0 \\
\varepsilon _{xy} & \varepsilon _{yy} & 0 \\
0 & 0 & \varepsilon _{zz}%
\end{array}%
\right)\label{strainf} \eequ with $\varepsilon _{zz}=-(\varepsilon
_{xx}+\varepsilon _{xx})$ expressed in an orthogonal coordinate
system ${Oxyz}$.

Assume that one end of the chain is fixed at origin $O$ while
another is denoted by $\textbf{R}_N$ before undergoing the
strains, where $N$ is the number of the segments of the chain. The
partition function of the chain can be calculated by path
integrals \cite{Dio}:
\[
Z=\int_{\mathbf{R}_{0}}^{\mathbf{R}_{N}}D[\mathbf{R}_{n}]\exp \left[ -\frac{3}{%
2L^{2}}\int_{0}^{N}dn\left( \frac{\partial \mathbf{R}_{n}}{\partial n}%
\right) ^{2}\right] =\sigma \exp \left[ -\frac{3(\mathbf{R}_{N}-\mathbf{R}%
_{0})^{2}}{2NL^{2}}\right],
\]
where $\sigma$ is a constant, and $L$ is the segment length. After
undergoing the strains, the partition function is changed to
\[
Z_{\epsilon }=\sigma \exp \left[ -\frac{3(\mathbf{R}_{N}-\mathbf{R}%
_{0})_{\epsilon }^{2}}{2NL^{2}}\right].
\]

Considering the relation $
(\mathbf{R}_{N}-\mathbf{R}_{0})_{\epsilon }^{2}=[(1+\epsilon )\cdot (\mathbf{%
R}_{N}-\mathbf{R}_{0})]^{2}$ and the distribution function of
end-to-end distance
$P(\mathbf{R}_{N}-\mathbf{R}_{0})=\frac{Z}{\int
d\mathbf{R}_{N}Z}$, we can obtain the free energy change as a
result of the strains: \beqn f_s&=&-k_{B}T\int d\mathbf{R}_{N}(\ln
Z_{\epsilon }-\ln
Z)P(\mathbf{R}_{N}-\mathbf{R}_{0})\nonumber\\&=&k_{B}T[(\varepsilon
_{xx}+\varepsilon _{yy})^{2}-(\varepsilon _{xx}\varepsilon
_{yy}-\varepsilon _{xy}^{2})].\label{gausschains} \eeqn

For the cell membrane with the cross-linking structure, we assume
that: (i) The membrane is a smooth surface and junction points
between protein chains are confined in the vicinity of the surface
and freely depart from it within the range of $\pm h/2$, where $h$
is the thickness of the membrane. (ii) There is no change of
volume occupied by the cross-linking structure on deformation. On
average, there are $\mathcal{N}$ protein chains per volume. (iii)
The protein chain can be regarded as a Gaussian chain with the
mean end-to-end distance much smaller than the dimension of the
cell membrane. (iv) The junction points move on deformation as if
they were embedded in an elastic continuum (Affine deformation
assumption). (v) The free energy of the cell membrane on
deformation is the sum of the free energies of closed lipid
bilayer and the cross-linking structure. The free energy of
cross-linking structure is the sum of the free energies of
individual protein chains.

If the cell membrane undergoing the small in-plane deformation
$\left(
\begin{array}{cc}
\varepsilon _{11} & \varepsilon _{12} \\
\varepsilon _{12} & \varepsilon _{22}
\end{array}
\right)$, where ``1" and ``2" represent two orthogonal directions
of the membrane surface, we can obtain the free energy of a
protein chain on deformation with the similar form of
(\ref{gausschains}) under above assumptions (i)$\sim$(iv). Using
above assumption (v), we have the free energy of the cell membrane
under the osmotic pressure $p$: \bequ
\mathcal{F}=\int_M(\mathcal{E}_d+\mathcal{E}_H)dA+p\int_VdV,
\label{freecellm}\eequ where $\mathcal{E}_H=(k_c/2)(2H+c_0)^2+\mu$
and $\mathcal{E}_d=(k_d/2)[(2J)^{2}-Q]$ with
$k_d=2\mathcal{N}hk_BT$, $2J=\varepsilon _{11}+\varepsilon _{22}$,
$Q=\varepsilon _{11}\varepsilon _{22}-\varepsilon _{12}^2$.

{\remk We do not write the term of $\bar{k}K$ in
{\rm(\ref{freecellm})} because $\int_M\bar{k}KdA$ is a constant
for closed surface {\rm (see \ref{appgbf})}.}

\subsection{Strain analysis}
If a point $\textbf{r}_0$ in a surface undergoing a displacement
$\textbf{u}$ to arrive at point $\textbf{r}$, we have $d\mathbf{u}
=d\mathbf{r}-d\mathbf{r}_{0}$ and $\delta _{i}d\mathbf{u} =\delta
_{i}d\mathbf{r}$ ($i=1,2,3$).

If denote $d\mathbf{r}=\omega _{1}\mathbf{e}_{1}+\omega
_{2}\mathbf{e}_{2}$ and $d\mathbf{u} =\mathbf{U}_{1}\omega
_{1}+\mathbf{U}_{2}\omega _{2}$ with $|\mathbf{U}_{1}| \ll
1,|\mathbf{U}_{2}|\ll 1$, we can define the strains \cite{wujk}:
\beqn
\varepsilon _{11} &=&\left[ \frac{d\mathbf{u}\cdot \mathbf{e}_{1}}{|d\mathbf{%
r}_{0}|}\right] _{\omega _{2}=0}\approx \mathbf{U}_{1}\cdot \mathbf{e}_{1},\label{epsl11} \\
\varepsilon _{22} &=&\left[ \frac{d\mathbf{u}\cdot \mathbf{e}_{2}}{|d\mathbf{%
r}_{0}|}\right] _{\omega _{1}=0}\approx \mathbf{U}_{2}\cdot \mathbf{e}_{2},\label{epsl22} \\
\varepsilon _{12} &=&\frac{1}{2}\left[ \left( \frac{d\mathbf{u}\cdot \mathbf{%
e}_{2}}{|d\mathbf{r}_{0}|}\right) _{\omega _{2}=0}+\left( \frac{d\mathbf{u}%
\cdot \mathbf{e}_{1}}{|d\mathbf{r}_{0}|}\right) _{\omega
_{1}=0}\right]
\approx \frac{1}{2}\left( \mathbf{U}_{1}\cdot \mathbf{e}_{2}+\mathbf{U}%
_{2}\cdot \mathbf{e}_{1}\right).\label{epsl12} \eeqn

Using $\delta _{i}d\mathbf{u} =\delta _{i}d\mathbf{r}$ and the
definitions of strains (\ref{epsl11})$\sim$(\ref{epsl12}), we can
obtain the variational relations:
\begin{eqnarray*}
\delta _{i}\varepsilon _{11}\omega _{1}\wedge \omega _{2}
&=&(1-\varepsilon _{11})\delta _{i}\omega _{1}\wedge\omega _{2}
-\mathbf{U}_{2}\cdot \mathbf{e}_{1}\delta _{i}\omega_{2}\wedge
\omega _{2}\\&&+\Omega _{i12}\mathbf{U}_{1}\cdot
\mathbf{e}_{2}\omega _{1}\wedge \omega _{2}+\Omega
_{i13}\mathbf{U}_{1}\cdot \mathbf{e}_{3}\omega _{1}\wedge \omega _{2}, \\
\delta _{i}\varepsilon _{12}\omega _{1}\wedge \omega _{2}
&=&\frac{1}{2}[(1-\varepsilon _{11})\omega _{1}\wedge \delta
_{i}\omega _{1}+(1-\varepsilon _{22})\delta _{i}\omega _{2}\wedge
\omega _{2}-\mathbf{U}
_{2}\cdot \mathbf{e}_{1}\omega _{1}\wedge \delta _{i}\omega _{2}\\&&-\mathbf{U}%
_{1}\cdot \mathbf{e}_{2}\delta _{i}\omega _{1}\wedge \omega
_{2}+\Omega _{i21}(\varepsilon _{11}-\varepsilon _{22})\omega
_{1}\wedge \omega _{2} \\
&&+\Omega _{i23}\mathbf{U}_{1}\cdot \mathbf{e}_{3}\omega
_{1}\wedge \omega _{2}+\Omega _{i13}\mathbf{U}_{2}\cdot
\mathbf{e}_{3}\omega _{1}\wedge
\omega _{2}], \\
\delta _{i}\varepsilon _{22}\omega _{1}\wedge \omega _{2}
&=&(1-\varepsilon _{22})\omega _{1}\wedge \delta _{i}\omega _{2}-\mathbf{U}%
_{1}\cdot \mathbf{e}_{2}\omega _{1}\wedge \delta _{i}\omega
_{1}\\&&+\Omega _{i21}\mathbf{U}_{2}\cdot \mathbf{e}_{1}\omega
_{1}\wedge \omega _{2}+\Omega _{i23}\mathbf{U}_{2}\cdot
\mathbf{e}_{3}\omega _{1}\wedge \omega _{2}.
\end{eqnarray*}
The leading terms of above relations are: \beqn \delta
_{i}\varepsilon _{11}\omega _{1}\wedge \omega _{2} &=&\delta
_{i}\omega _{1}\wedge \omega _{2},\label{epsli11} \\
\delta _{i}\varepsilon _{12}\omega _{1}\wedge \omega _{2} &=&\frac{1}{%
2}[\omega _{1}\wedge \delta _{i}\omega _{1}+\delta _{i}\omega
_{2}\wedge
\omega _{2}],\label{epsli12} \\
\delta _{i}\varepsilon _{22}\omega _{1}\wedge \omega _{2}
&=&\omega _{1}\wedge \delta _{i}\omega _{2}.\label{epsli22} \eeqn
Thus, \beqn \delta _{i}(2J)\omega _{1}\wedge \omega _{2} &=&\delta
_{i}(\varepsilon _{11}+\varepsilon _{22})\omega _{1}\wedge \omega
_{2}=\delta _{i}\omega _{1}\wedge \omega _{2}+\omega _{1}\wedge
\delta _{i}\omega _{2},\label{deltai2J}
\\
\delta _{i}Q\omega _{1}\wedge \omega _{2} &=&\delta
_{i}(\varepsilon
_{11}\varepsilon _{22}-\varepsilon _{12}^{2})\omega _{1}\wedge \omega _{2}\nonumber \\
&=&(\varepsilon _{11}\omega _{1}+\varepsilon _{12}\omega
_{2})\wedge \delta _{i}\omega _{2}-(\varepsilon _{12}\omega
_{1}+\varepsilon _{22}\omega _{2})\wedge \delta _{i}\omega
_{1}.\label{deltaiQ} \eeqn

Considering equations (\ref{omvaratione11})$\sim$(\ref{domega3}),
(\ref{deltai2J}) and (\ref{deltaiQ}), we have \beqn \delta
_{1}(2J)\omega _{1}\wedge \omega _{2} &=&d(\Omega
_{1}\omega _{2}),\label{delta12J} \\
\delta _{1}Q\omega _{1}\wedge \omega _{2} &=&(\varepsilon
_{11}d\omega _{2}-\varepsilon _{12}d\omega _{1})\Omega
_{1}-(\varepsilon _{12}\omega _{1}+\varepsilon _{22}\omega
_{2})\wedge d\Omega _{1}; \label{delta1Q}\eeqn \beqn \delta
_{2}(2J)\omega _{1}\wedge \omega _{2} &=&-d(\Omega
_{2}\omega _{1}),\label{delta22J} \\
\delta _{2}Q\omega _{1}\wedge \omega _{2} &=&(\varepsilon
_{11}\omega _{1}+\varepsilon _{12}\omega _{2})\wedge d\Omega
_{2}+\Omega _{2}(\varepsilon _{12}d\omega _{2}-\varepsilon
_{22}d\omega _{1});\label{delta2Q} \eeqn \beqn \delta
_{3}(2J)\omega _{1}\wedge \omega _{2} &=&-2H\Omega
_{3}dA,\label{delta32J} \\
\delta _{3}Q\omega _{1}\wedge \omega _{2} &=&[-2H(2J)+a\varepsilon
_{11}+2b\varepsilon _{12}+c\varepsilon _{22}]\Omega
_{3}dA.\label{delta3Q} \eeqn

\subsection{Shape equation and in-plane strain equations of cell membranes}
To obtain the shape equation and in-plane strain equations of cell
membranes, we must take the first order variation of the
functional (\ref{freecellm}). Denote
$\mathcal{F}_d=\int_M\mathcal{E}_ddA$ and
$\mathcal{F}_{cp}=\int_M\mathcal{E}_HdA+p\int_VdV$.

From (\ref{delta12J})$\sim$(\ref{delta3Q}), we can calculate that:
\beqn
\delta _{1}\mathcal{F}_{d} &=&\int_{M}\left[ \frac{\partial \mathcal{E}_{d}}{%
\partial (2J)}\delta _{1}(2J)dA+\frac{\partial \mathcal{E}_{d}}{\partial Q}%
\delta _{1}QdA+\mathcal{E}_{d}(2J,Q)\delta _{1}dA\right]\nonumber \\
&=&-\int_{M}d\left[ \frac{\partial \mathcal{E}_{d}}{\partial (2J)}+\mathcal{E%
}_{d}(2J,Q)\right] \wedge \omega _{2}\Omega
_{1}+\int_{M}\frac{\partial \mathcal{E}_{d}}{\partial
Q}(\varepsilon _{11}d\omega _{2}-\varepsilon
_{12}d\omega _{1})\Omega _{1}\nonumber \\
&&-\int_{M}\Omega _{1}d\left[ (\varepsilon _{12}\omega
_{1}+\varepsilon _{22}\omega _{2})\frac{\partial
\mathcal{E}_{d}}{\partial Q}\right], \eeqn \beqn
\delta _{2}\mathcal{F}_{d} &=&\int_{M}\left[ \frac{\partial \mathcal{E}_{d}}{%
\partial (2J)}\delta _{2}(2J)dA+\frac{\partial \mathcal{E}_{d}}{\partial Q}%
\delta _{2}QdA+\mathcal{E}_{d}(2J,Q)\delta _{2}dA\right]\nonumber \\
&=&\int_{M}d\left[ \frac{\partial \mathcal{E}_{d}}{\partial (2J)}+\mathcal{E}%
_{d}(2J,Q)\right] \wedge \omega _{1}\Omega
_{2}+\int_{M}\frac{\partial \mathcal{E}_{d}}{\partial
Q}(\varepsilon _{12}d\omega _{2}-\varepsilon
_{22}d\omega _{1})\Omega _{2}\nonumber \\
&&+\int_{M}\Omega _{2}d\left[ \frac{\partial \mathcal{E}_{d}}{\partial Q}%
[(\varepsilon _{11}\omega _{1}+\varepsilon _{12}\omega
_{2})\right], \eeqn \beqn
\delta _{3}\mathcal{F}_{d} &=&\int_{M}\left[ \frac{\partial \mathcal{E}_{d}}{%
\partial (2J)}\delta _{3}(2J)dA+\frac{\partial \mathcal{E}_{d}}{\partial Q}%
\delta _{3}QdA+\mathcal{E}_{d}(2J,Q)\delta _{3}dA\right]\nonumber \\
&=&\int_{M}\frac{\partial \mathcal{E}_{d}}{\partial
Q}[a\varepsilon _{11}+2b\varepsilon _{12}+c\varepsilon _{22}]
\Omega _{3}dA\nonumber\\&&-\int_{M}2H\left[ \frac{\partial
\mathcal{E}_{d}}{\partial (2J)}+\mathcal{E}_{d}+(2J)\frac{\partial \mathcal{E%
}_{d}}{\partial Q}\right] \Omega _{3}dA.\label{delta3fd} \eeqn

Otherwise, section \ref{close-bilayer} tells us:
\begin{eqnarray*}
\delta _{1}\mathcal{F}_{cp} &=&\delta _{2}\mathcal{F}_{cp}=0, \\
\delta _{3}\mathcal{F}_{cp} &=&\int_{M}\left[ (\nabla ^{2}+4H^{2}-2K)\frac{%
\partial \mathcal{E}_{H}}{\partial (2H)}-2H\mathcal{E}_{H}+p\right] \Omega
_{3}dA.
\end{eqnarray*}

Therefore, $\delta _{i}\mathcal{F}=\delta _{i}\mathcal{F}_d+\delta
_{i}\mathcal{F}_{cp}=0$ gives: \beqn
&&-d\left[ \frac{\partial \mathcal{E}_{d}}{\partial (2J)}+\mathcal{E}_{d}%
\right] \wedge \omega _{2}+\frac{\partial \mathcal{E}_{d}}{\partial Q}%
(\varepsilon _{11}d\omega _{2}-\varepsilon _{12}d\omega
_{1})\nonumber\\&&\qquad-d\left[ (\varepsilon _{12}\omega
_{1}+\varepsilon _{22}\omega _{2})\frac{\partial
\mathcal{E}_{d}}{\partial Q}\right]=0, \\
&&d\left[ \frac{\partial \mathcal{E}_{d}}{\partial
(2J)}+\mathcal{E}_{d}\right] \wedge \omega _{1}+\frac{\partial
\mathcal{E}_{d}}{\partial Q}(\varepsilon _{12}d\omega
_{2}-\varepsilon _{22}d\omega _{1})\nonumber\\&&\qquad+d\left[
\frac{\partial \mathcal{E}_{d}}{\partial Q}(\varepsilon
_{11}\omega _{1}+\varepsilon
_{12}\omega _{2})\right] =0, \\
&&(\nabla ^{2}+4H^{2}-2K)\frac{\partial \mathcal{E}_{H}}{\partial (2H)}-2H%
\left[ \mathcal{E}_{H}+\frac{\partial \mathcal{E}_{d}}{\partial (2J)}+%
\mathcal{E}_{d}+(2J)\frac{\partial \mathcal{E}_{d}}{\partial Q}\right]\nonumber\\
&&\qquad+p+\frac{\partial \mathcal{E}_{d}}{\partial
Q}[a\varepsilon _{11}+2b\varepsilon _{12}+c\varepsilon _{22}] =0.
\eeqn Substituting $\mathcal{E}_{H}
=\frac{k_{c}}{2}(2H+c_{0})^{2}+\mu$ and $\mathcal{E}_{d}
=\frac{k_{d}}{2}[(2J)^{2}-Q]$ into above three equations, we
obtain: \beqn &&k_d[-d(2J)\wedge \omega
_{2}-\frac{1}{2}(\varepsilon _{11}d\omega _{2}-\varepsilon
_{12}d\omega _{1})+\frac{1}{2}d(\varepsilon _{12}\omega
_{1}+\varepsilon _{22}\omega _{2})] =0,\label{shapecm1} \\
&&k_d[d(2J)\wedge \omega _{1}-\frac{1}{2}(\varepsilon _{12}d\omega
_{2}-\varepsilon _{22}d\omega _{1})-\frac{1}{2}d(\varepsilon
_{11}\omega
_{1}+\varepsilon _{12}\omega _{2})] =0, \label{shapecm2}\\
&&p-2H(\mu +k_{d}J)+k_{c}(2H+c_{0})(2H^{2}-c_{0}H-2K)+k_{c}\nabla
^{2}(2H)\nonumber\\&&\qquad-\frac{k_{d}}{2}(a\varepsilon
_{11}+2b\varepsilon _{12}+c\varepsilon _{22}) =0.
\label{shapecm3}\eeqn (\ref{shapecm1}) and (\ref{shapecm2}) are
called in-plane strain equations of the cell membrane, while
(\ref{shapecm3}) is the shape equation.

{\remk The higher order terms of $\varepsilon _{ij}$ ($i,j=1,2$)
are neglected in above three equations.}

Obviously, if $k_d=0$, then (\ref{shapecm1}) and (\ref{shapecm2})
are two identities. Moreover (\ref{shapecm3}) degenerates into
shape equation (\ref{shape-closed}) of closed lipid bilayers in
this case. Otherwise, for small strain, (\ref{shapecm3}) is very
close to (\ref{shape-closed}), which may suggest that
cross-linking structures have small effects on the shape of lipid
bilayers.

It is not hard to verify that $\varepsilon_{12}=0$,
$\varepsilon_{11}=\varepsilon_{22}=\varepsilon$ satisfy
(\ref{shapecm1}) and (\ref{shapecm2}) if $\varepsilon$ being a
constant. In this case, the sphere with radius $R$ is the solution
of (\ref{shapecm3}) if it satisfies \bequ pR^{2}+(2\mu
+3k_{d}\varepsilon )R+k_{c}c_{0}(c_{0}R-2)
=0.\label{sphericalcell}\eequ

\subsection{Mechanical stabilities of spherical cell membranes}
To discuss the stabilities of spherical cell membranes, we must
discuss the second order variations of the functional
$\mathcal{F}$. In mathematical point of view presented in section
\ref{Math-pre}, we must calculate $\delta_i\delta_j\mathcal{F}$
($i,j=1,2,3$). But in physical and symmetric point of view, we
just need to calculate $\delta_3^2\mathcal{F}$ because we can
expect that the perturbations along the normal are primary to the
instabilities of spherical membranes under the osmotic pressure
which is perpendicular to the sphere surfaces.

If we taking $\mathcal{E}_{H} =\frac{k_{c}}{2}(2H+c_{0})^{2}+\mu$
and $\mathcal{E}_{d} =\frac{k_{d}}{2}[(2J)^{2}-Q]$, the leading
term of (\ref{delta3fd}) is \bequ \delta
_{3}\mathcal{F}_{d}=-\frac{k_{d}}{2}\int_{M}\left[ (a\varepsilon
_{11}+2b\varepsilon _{12}+c\varepsilon _{22})+2H(2J)\right] \Omega
_{3}dA.\label{delta3fdsim} \eequ

Using Lemmas \ref{delta3da}, \ref{delta32H},
Eqs.(\ref{epsli11})$\sim$(\ref{epsli22}), we can obtain \beqn
\delta _{3}^{2}\mathcal{F}_{d} &=&-\frac{k_{d}}{2}\int_{M}\left[
(\delta _{3}a\varepsilon _{11}+2\delta _{3}b\varepsilon
_{12}+\delta
_{3}c\varepsilon _{22})+\delta _{3}(2H)(2J)\right] \Omega _{3}dA\nonumber \\
&&-\frac{k_{d}}{2}\int_{M}\left[ (a\delta _{3}\varepsilon
_{11}+2b\delta _{3}\varepsilon _{12}+c\delta _{3}\varepsilon
_{22})+2H\delta _{3}(2J)\right]
\Omega _{3}dA \nonumber\\
&&-\frac{k_{d}}{2}\int_{M}\left[ (a\varepsilon _{11}+2b\varepsilon
_{12}+c\varepsilon _{22})+2H(2J)\right] \Omega _{3}\delta _{3}dA \nonumber\\
&=&k_{d}\int_{M}\frac{3(1+\varepsilon )}{R^{2}}\Omega _{3}^{2}dA-\frac{%
3k_{d}\varepsilon }{2}\int_{M}\Omega _{3}\nabla ^{2}\Omega
_{3}dA,\label{secdelta3Fd} \eeqn for the spherical cell membrane
with radius $R$ and strain $\varepsilon$.

Otherwise, (\ref{secondvarnoK}) suggests that \beqn \delta
_{3}^{2}\mathcal{F}_{cp} &=&\int_{M}\Omega
_{3}^{2}\{k_{c}c_{0}^{2}/R^{2}+2\mu /R^{2}+2p/R\}dA \nonumber\\
&&+\int_{M}\Omega _{3}\nabla ^{2}\Omega
_{3}\{2k_{c}c_{0}/R+2k_{c}/R^{2}-\mu
-k_{c}c_{0}^{2}/2\}dA \nonumber\\
&&+\int_{M}k_{c}(\nabla ^{2}\Omega _{3})^{2}dA. \eeqn

Therefore \beqn \delta _{3}^{2}\mathcal{F} &=&\delta
_{3}^{2}\mathcal{F}_d+\delta _{3}^{2}\mathcal{F}_{cp}\nonumber\\
&=&\int_{M}\Omega _{3}^{2}\{3k_{d}/R^{2}+(3k_{d}\varepsilon
+k_{c}c_{0}^{2}+2\mu )/R^{2}+2p/R\}dA\nonumber \\
&&+\int_{M}\Omega _{3}\nabla ^{2}\Omega
_{3}\{2k_{c}c_{0}/R+2k_{c}/R^{2}-(3k_{d}\varepsilon
+k_{c}c_{0}^{2}+2\mu
)/2\}dA \nonumber\\
&&+\int_{M}k_{c}(\nabla ^{2}\Omega _{3})^{2}dA.\label{secdeltaFcm}
\eeqn

If considering (\ref{sphericalcell}) and expanding $\Omega _{3}$
as (\ref{harmonicf}), we have
\begin{eqnarray*}
\delta _{3}^{2}\mathcal{F} &=&\int_{M}\Omega
_{3}^{2}\{3k_{d}/R^{2}+(2k_{c}c_{0}/R^{3})+p/R\}dA \\
&&+\int_{M}\Omega _{3}\nabla ^{2}\Omega
_{3}\{k_{c}c_{0}/R+2k_{c}/R^{2}+pR/2\}dA+\int_{M}k_{c}(\nabla
^{2}\Omega _{3})^{2}dA\\
&=&\sum_{l,m}|a_{lm}|^{2}%
\{3k_{d}+[l(l+1)-2][l(l+1)k_{c}/R^{2}-k_{c}c_{0}/R-pR/2]\}.
\end{eqnarray*}

The zero point of the coefficient of $|a_{lm}|^2$ in above
expression is \bequ p_{l}
=\frac{6k_{d}}{[l(l+1)-2]R}+\frac{2k_{c}[l(l+1)-c_{0}R]}{R^{3}}\quad(l=2,3,\cdots).\label{criticalpsl}
\eequ Obviously, on the one hand, if $k_d=0$, (\ref{criticalpsl})
is degenerated into (\ref{criticalps}) with $l\geq 2$. On the
other hand, if $k_d>0$, we must take the minimum of
(\ref{criticalpsl}) to obtain the critical pressure.

If let $\xi=l(l+1)\geq 6$, we have \beqn p(\xi)
&=&\frac{6k_{d}}{(\xi-2)R}+\frac{2k_{c}(\xi-c_{0}R)}{R^{3}},\label{pxi}\\
\frac{dp}{d\xi } &=&-\frac{6k_{d}/R}{(\xi -2)^{2}}+\frac{2k_{c}}{R^{3}},\label{dpxi} \\
\frac{d^{2}p}{d\xi ^{2}} &=&\frac{12k_{d}/R}{(\xi -2)^{3}}>0.
\label{d2pxi}\eeqn $dp/d\xi=0$ and $\xi\geq 6$ imply $\xi
=2+R\sqrt{3k_{d}/k_{c}}$ which is valid only if
$3k_{d}R^{2}>16k_{c}$. Therefore, the critical pressure is: \bequ
p_{c}=\min \{p_{l}\}=\left\{
\begin{array}{c}
\frac{3k_{d}}{2R}+\frac{2k_{c}(6-c_{0}R)}{R^{3}}<\frac{2k_{c}(10-c_{0}R)}{%
R^{3}}\quad (3k_{d}R^{2}<16k_{c}), \\
\frac{4\sqrt{3k_{d}k_{c}}}{R^{2}}+\frac{2k_{c}}{R^{3}}(2-c_{0}R)\quad
(3k_{d}R^{2}>16k_{c}).
\end{array}%
\right. \label{criticalpcm}\eequ

Eq.(\ref{criticalpcm}) includes the classical result for stability
of elastic shell. The critical pressure for classical spherical
shell is $p_c\propto Yh^2/R^2$ \cite{landau,Pogorelov}, where $Y$
is the Young's modulus of the shell. If taking $c_0=0$,
$k_d\propto Yh$, $k_c\propto Yh^3$ and $R\gg h$, our result
(\ref{criticalpcm}) also gives $p_c\propto Yh^2/R^2$. As far as we
know, this is the first time to obtain the critical pressure for
spherical shell through the second order variation of free energy
without any assumption to the shape of its losing the stability
(cf. Ref.\cite{landau,Pogorelov}).

Otherwise, if we take the typical parameters of cell membranes as
$k_c\sim 20k_BT$ \cite{Duwe,Mutz2}, $k_d\sim 2.4\mu N/m$
\cite{Lenormand}, $h\sim 4nm$, $R\sim 1\mu m$, $c_0R\sim 1$, we
obtain $p_c\sim4$ Pa from (\ref{criticalpcm}), which is much
larger than $p_c\sim 0.2 $Pa without considering $k_d$ induced by
the cross-linking structures. Therefore, cross-linking structures
greatly enhance the mechanical stabilities of cell membranes.

\section{Conclusion\label{conclusion}}
In above discussion, we deal with variational problems on closed
and open surfaces by using exterior differential forms. We obtain
the shape equation of closed lipid bilayers, the shape equation
and boundary conditions of open lipid bilayers and two-component
lipid bilayers, and the shape equation and in-plane stain
equations of cell membranes with cross-linking protein structures.
Furthermore, we discuss the mechanical stabilities of spherical
lipid bilayers and cell membranes.

Some new results are obtained as follows:

(i) The fundamental variational equations in a surface: Eqs.
(\ref{omvaratione11}) $\sim$ (\ref{detaomegaij}).

(ii) The general expressions of the second order variation of
 the free energy for closed lipid bilayers: theorem \ref{thvar2Fc}
 and Eq. (\ref{secondvarnoK}).

(iii) The general shape equation and boundary conditions of open
lipid bilayers and two-component lipid bilayers: Eqs.
(\ref{euleropen1}) $\sim$(\ref{euleropen4}) and
(\ref{shape-twocom})$\sim$(\ref{boundtwocom3}).

(iv) The free energy (\ref{freecellm}), shape equation and strain
equations (\ref{shapecm1})$\sim$(\ref{shapecm3}) of the cell
membranes with cross-linking protein structures.

(v) The critical pressure (\ref{criticalpcm}) of losing
stabilities for spherical cell membranes. It includes the critical
pressures not only for closed lipid bilayers, but also for the
classic solid shells. Otherwise it suggests that cross-linking
protein structures can enhance the stabilities of cell membranes.

In the future, we will devote ourselves to applying above results
to explain the shapes of open lipid bilayers found by Saitoh
\textit{et al.}, and predict new shapes of multi-component lipid
bilayers and cell membranes. Moreover, We will discuss whether and
how the in-plane modes affect the instability of cell membranes
although we believe they have no qualitative effect on our results
in section 5.4.

\ack{We are grateful to Prof. H W Peng for his useful discussions
and to Prof. S S Chern for his advice that we should notice the
work by Griffiths \textit{et al.} Thank Prof. J Guven and G
Landolfi for their friendly email discussions. Thank Prof. J Hu,
Dr. W Zhao and R An for their kind helps.}

\appendix
\section{Exterior differential forms and Stokes' theorem \label{appdiff}}
A manifold can be roughly regarded as a multi-dimensional surface.
In the neighborhood of every point, we can construct the local
coordinates $(u^1,u^2,\cdots,u^m)$, where $m$ is the dimension of
the surface. In this paper we just consider smooth, orientable
manifolds and smooth functions.

We call the function $f(u^1,u^2,\cdots,u^m)$ 0-form and
$a_i(u^1,u^2,\cdots,u^m)du^i$ 1-form, where Einstein summation
rule is used and it is also used in the following contents. The
$r$-form ($r\leq m$) is defined as $a_{i_1i_2\cdots
i_r}du^{i_1}\wedge du^{i_2}\wedge\cdots\wedge du^{i_r}$, where the
exterior production ``$\wedge$" satisfies $du^i\wedge
du^j=-du^j\wedge du^i$. Denote $\Lambda^r=\{\textrm{all
}r\textrm{-forms}\}$, $(r=0,1,2,\cdots,m)$.

\noindent\textbf{Definition} A linear operator
$d:\Lambda^r\rightarrow\Lambda^{r+1}$ is called the exterior
differential operator if it satisfies:

{\rm (i)} \ For function $f(u^1,u^2,\cdots,u^m)$,
$df=\frac{\partial f}{\partial u^i}du^i$ is an ordinary
differential;

{\rm (ii)} \ $dd=0$;

{\rm (iii)} \ $\forall\omega_1\in \Lambda^r$ \textrm{and}
$\forall\omega_2\in\Lambda^k$,
$d(\omega_1\wedge\omega_2)=d\omega_1\wedge\omega_2+(-1)^r
\omega_1\wedge d\omega_2$.\\
\textbf{Stokes theorem} If $\omega$ is an $(m-1)$-form with
compact support set on $M$, and $\mathcal{D}$ is a domain with
boundary $\partial \mathcal{D}$ in $M$, then
\begin{equation}\label{stokes}
\int_{\mathcal{D}} d\omega=\int_{\partial\mathcal{D}}\omega.
\end{equation}

\section{Curves in a surface \label{appcurve}}
If a curve passes through $P$ in the surface, we construct a Frenet frame $\{%
\mathbf{T},\mathbf{N},\mathbf{B}\}$ such that $\mathbf{T}$,
$\mathbf{N}$ and $\mathbf{B}$ are the tangent, normal and binormal
vectors of the curve, respectively. Denote $\theta$ the angle
between $\mathbf{e}_{1}$ and $\mathbf{T}$. Set
$\mathbf{M}=\mathbf{e}_{3}\times \mathbf{T}$. Thus we have
\[\left\{
\begin{array}{l}
\mathbf{T}=\mathbf{e}_{1}\cos \theta +\mathbf{e}_{2}\sin \theta , \\
\mathbf{M}=-\mathbf{e}_{1}\sin \theta +\mathbf{e}_{2}\cos \theta .
\end{array}
\right. \] It is not hard to calculate
\[
d\mathbf{T} =(d\theta +\omega
_{12})\mathbf{M}+\mathbf{e}_{3}(\omega _{13}\cos \theta +\omega
_{23}\sin \theta).
\]

Frenet Formulas tell us $d\mathbf{T}/ds=\kappa \mathbf{N}$.
Therefore, we have the geodesic curvature, the geodesic torsion,
and the normal curvature of the curve: \beqn
k_{g}=\kappa \mathbf{N}\cdot \mathbf{M}=(d\mathbf{T}/ds)\cdot \mathbf{M}%
=(d\theta +\omega _{12})/ds,\label{geodisicc}\\
\tau _{g} =-(d\mathbf{e}_{3}/ds)\cdot \mathbf{M}=[(b\omega
_{1}+c\omega _{2})\cos \theta-(a\omega _{1}+b\omega _{2})(\sin
\theta)]/ds,\nonumber\\
k_{n}=II/I=(a\omega _{1}^{2}+2b\omega _{1}\omega _{2}+c\omega
_{2}^{2})/(\omega _{1}^{2}+\omega _{2}^{2}). \nonumber\eeqn

If the curve along $\mathbf{e}_{1}$ such that $\theta =0$, we have
$ds=\omega _{1},\omega _{2}=0$ and \bequ k_{g} =\omega
_{12}/\omega _{1},\quad\tau_{g}=b,\quad  \textrm{and}\quad k_{n}
=a.\label{curve-surfaceapp}\eequ

\section{Gauss-Bonnet formula \label{appgbf}}
Using (\ref{domgaij}), (\ref{omega13}) and (\ref{gassianK}), we
have \bequ\label{Egregium}
d\omega_{12}=-K\omega_{1}\wedge\omega_{2}.\eequ This formula was
called \emph{Theorem Egregium} by Gauss. From \emph{Theorem
Egregium} and (\ref{geodisicc}), we can derive Gauss-Bonnet
formula: \bequ\label{Gauss-Bonnet} \int_M K dA+\int_C k_g
ds=2\pi\chi(M),\eequ where $\chi(M)$ is the characteristic number
of smooth surface $M$ with smooth edge $C$. $\chi(M)=1$ for a
simple surface with an edge. For a closed surface, we have
\bequ\label{Gaussclosed} \int_M K dA=2\pi\chi(M).\eequ

\section{The tensor expressions of $\nabla$, $\bar{\nabla}$,
$\tilde{\nabla}$, $\nabla^2$, $\nabla\cdot\bar{\nabla}$, and
$\nabla\cdot\tilde{\nabla}$ \label{app2}}

At every point $\mathbf{r}$ in the surface, we can take local
coordinates $(u^1,u^2)$ where the first and the second fundamental
form are denoted by $I=g_{ij}du^idu^j$ and $II=L_{ij}du^idu^j$,
respectively. Let $(g^{ij})=(g_{ij})^{-1}$,
$(L^{ij})=(L_{ij})^{-1}$ and $\mathbf{r}_i=\partial
\mathbf{r}/\partial u^i$, thus we have
\begin{eqnarray*}
\nabla  =g^{ij}\mathbf{r}_{i}\frac{\partial }{\partial u^{j}}, \\
\bar{\nabla} =\mathbf{r}_{i}(2Hg^{ij}-KL^{ij})\frac{\partial
}{\partial
u^{j}}, \\
\tilde{\nabla} =KL^{ij}\mathbf{r}_{i}\frac{\partial }{\partial
u^{j}},
\\
\nabla ^{2} =\frac{1}{\sqrt{g}}\frac{\partial }{\partial
u^{i}}\left(
\sqrt{g}g^{ij}\frac{\partial }{\partial u^{j}}\right),  \\
\nabla \cdot \bar{\nabla} =\frac{1}{\sqrt{g}}\frac{\partial
}{\partial
u^{i}}\left[ \sqrt{g}(2Hg^{ij}-KL^{ij})\frac{\partial }{\partial u^{j}}%
\right],  \\
\nabla \cdot \tilde{\nabla} =\frac{1}{\sqrt{g}}\frac{\partial
}{\partial u^{i}}\left( \sqrt{g}KL^{ij}\frac{\partial }{\partial
u^{j}}\right).
\end{eqnarray*}

As an example, We will prove the last one of above expressions.\\
{\prf If taking the orthogonal local coordinates, we have $
I=g_{11}(du^{1})^{2}+g_{22}(du^{2})^{2}=\omega_1^2+\omega_2^2$,
which implies $\omega _{1}=\sqrt{g_{11}}du^{1}$ and $\omega
_{2}=\sqrt{g_{22}}du^{2}$. For function $f$, on the one hand, we
have $df(u^{1},u^{2}) =f_{1}\omega _{1}+f_{2}\omega _{2}
=f_{1}\sqrt{g_{11}}du^{1}+f_{2}\sqrt{g_{22}}du^{2}$, on the other
hand, we have
$df=\frac{\partial f}{\partial u^{1}}du^{1}+\frac{\partial f}{\partial u^{2}}%
du^{2}$. Therefore, $ f_{1} =\frac{1}{\sqrt{g_{11}}}\frac{\partial
f}{\partial u^{1}},\quad f_{2}=
\frac{1}{\sqrt{g_{22}}}\frac{\partial f}{\partial u^{2}}$.

The second fundamental form $II=a\omega _{1}^{2}+2b\omega
_{1}\omega _{2}+c\omega _{1}^{2}=L_{ij}du^{i}du^{j}$ implies $a
=L_{11}/g_{11},\quad b=L_{12}/\sqrt{g},\quad c=L_{22}/g_{22}$.
Thus $K=ac-b^{2}=(L_{11}L_{22}-L_{12}^{2})/g$, and
\begin{eqnarray*}
L^{11} &=&\frac{L_{22}}{L_{11}L_{22}-L_{12}^{2}}\Rightarrow
L_{22}=gKL^{11};
\\
L^{12} &=&-\frac{L_{12}}{L_{11}L_{22}-L_{12}^{2}}\Rightarrow
L_{12}=-gKL^{12};
\\
L^{22} &=&\frac{L_{11}}{L_{11}L_{22}-L_{12}^{2}}\Rightarrow
L_{11}=gKL^{22}.
\end{eqnarray*}

Moreover, we have
\begin{eqnarray*}
\tilde{\ast}\tilde{d}f &=&-f_{2}\omega _{13}+f_{1}\omega _{23}
=-f_{2}(a\omega_1+b\omega_2)+f_2(b\omega_1+c\omega_2) \\
&=&\frac{1}{\sqrt{g}}\left(L_{12}\frac{\partial f}{\partial u^{1}}-L_{11}\frac{%
\partial f}{\partial u^{2}}\right)du^{1}+\frac{1}{\sqrt{g}}\left(L_{22}\frac{\partial f%
}{\partial u^{1}}-L_{12}\frac{\partial f}{\partial u^{2}}\right)du^{2};\\
d\tilde{\ast}\tilde{d}f
&=&\left\{\frac{\partial}{\partial u^{1}}\left[\sqrt{g}K\left(L^{11}\frac{\partial f}{%
\partial u^{1}}+L^{12}\frac{\partial f}{\partial u^{2}}\right)\right]\right.\\
&& \quad +\left.\frac{\partial }{%
\partial u^{2}}\left[\sqrt{g}K\left(L^{12}\frac{\partial f}{\partial u^{1}}+L^{22}%
\frac{\partial f}{\partial
u^{2}}\right)\right]\right\}du^{1}\wedge du^{2}.
\end{eqnarray*}
Therefore, $
\nabla\cdot\tilde{\nabla}f=\frac{d\tilde{\ast}\tilde{d}f}{\omega
_{1}\wedge \omega _{2}} =\frac{1}{\sqrt{g}}\frac{\partial
}{\partial u^{i}} \left(\sqrt{g}KL^{ij}\frac{\partial f}{\partial
u^{j}}\right)$. \P}

\end{document}